\documentclass[twocolumn,aps]{revtex4-1}
\usepackage[section] {placeins}
\usepackage{epsfig}
\newcommand{\beq}{\begin{eqnarray}}
\newcommand{\eeq}{\end{eqnarray}}
\begin{document}
\title{ Simple and Accurate Oscillation Probabilities for Three Coupled Neutrinos Propagating in Matter}

\author{
Mikkel B. Johnson \\
Los Alamos National Laboratory, Los Alamos, NM 87545 \\
Leonard S. Kisslinger\\
Department of Physics, Carnegie-Mellon University, Pittsburgh, PA 15213\\
}
\begin{abstract}

Within a conventional Hamiltonian description, we find accurate closed-form expressions for the  oscillation probabilities  of three coupled neutrinos propagating in matter. Subtle cancelations that occur in coefficients of our formulation are avoided for all transitions  $\nu_a \to \nu_b$ by transforming to a different set of coefficients presented in an appendix of this paper.  The neutrino mass eigenvalues are easily obtained numerically as the solution of a cubic equation.  Our methods are illustrated for flavor-changing transitions in the $(\nu_e,\nu_\mu)$ sector.  The resulting analytic  expressions oscillation probabilities, which are  particularly simple, are also  accurate to a few percent over all regions of interest at present and the envisioned future neutrino facilities. While somewhat  less accurate than numerical simulations,  our approximate expressions are sufficiently accuracy to  obviate the need for exact  computer simulations in many circumstances.  
 
\end{abstract}

\maketitle

\noindent

PACS Indices:  11.3.Er,14.60.Lm,13.15.+g 

\vspace{1mm}

\noindent

Keywords:

\vspace{1mm}

\noindent

\section{Introduction}

In this paper, we develop methods leading to approximate analytical expressions for the oscillation probability from the exact, closed-form results in Ref.~\cite{jhk0}.  These results are based on a conventional Hamiltonian and the Standard Neutrino Model~\cite{ISS}  and lead to expressions for the oscillation probability  for all transitions  $\nu_a \to \nu_b$.  

We include  the interaction of neutrinos with electrons shown to be essential by Wolfenstein in 1978~\cite{wolf} and later identified by Mikheyev and Smirnov~\cite{SMAS} as a likely explanation of the deficit of solar neutrinos discovered experimentally by Davis~\cite{Davis,Clev}. 
 
Subsequent to the work of Wolfenstein,  neutrino oscillations  in matter have been  commonly explored using exact computer simulations.  These methods have advantages for  density profiles having a spatial dependence, are  discussed in  Ref.~\cite{MS} and references therein. 

Comparisons of Ref.~\cite{jhk0} to currently available analytical  results~\cite{f,ahlo,JO} are found in Ref.~\cite{jhk1}, where it is established that  these  results~\cite{f,ahlo,JO}   are reliable over rather narrow regions of energy, density, and baseline.  Expressions found here do not suffer from these limitations. 

In the present paper we use the published formulation of Ref.~\cite{jhk0}. Because we rely heavily on all results found in Ref.~\cite{jhk1}, the reader will find many of the important results of the unpublished  Ref~\cite{jhk1} repeated here.

Our formulation~\cite{jhk0} is  introduced in  Sect.~\ref{ND} with details relegated to  appendices.   The oscillation probability simplifies in part  because many of the terms appearing in the exact coefficients  $w^{(ab)}_{i,p}$ of our approach~\cite{jhk0} are quite small  [$O(\alpha^n)$, $n \geq 3$] and can therefore be eliminated.
 
The expressions we find for the oscillation probabilities are built on coefficients $w^{(ab)}_{i,p^-}$ that are devoid of subtle cancelations occurring in the coefficients $w^{(ab)}_{i,p}$ of  Ref.~\cite{jhk0}.   The nature of these cancelations is discussed in Sect.~\ref{simplifyOP}, and their source is identified in Sect.~\ref{f2ltw}.  The cancellations are avoided by making a transformation that leaves the partial oscillation probabilities invariant.  The resulting  coefficients $w^{(ab)}_{i,p^-}$ are essential for obtaining simple expressions for the observable oscillation probabilities accurate to $O(\alpha^2)$ [about 1\%].  The coefficients $w^{(ab)}_{i,p^-}$ are tabulated  in Appendix~\ref{a:TCoefw} of the present paper for all transitions  $\nu_a \to \nu_b$.

Observable oscillation probabilities accurate to $O(\alpha^2)$ are easily  found  using $w^{(ab)}_{i,p^-}$ and eigenvalues  obtained by numerically  by solving a cubic equation. However, the emphasis of this paper is on developing  methods for obtaining  simple and accurate  ${\it analytical}$ expressions for the oscillation probabilities.  These methods are based on the same coefficients $w^{(ab)}_{i,p^-}$.

Application of these methods is illustrated for flavor-changing transitions in the $(\nu_e,\nu_\mu)$ sector. The  oscillation probabilities so obtained  are   accurate  to a few percent over all regions of interest at present and envisioned future neutrino facilities.  While somewhat  less accurate than numerical simulations,  our approximate expressions are sufficiently accurate to  obviate the need for exact  computer simulations in many circumstances.  

Our results rely on neutrino mass eigenvalues  evaluated using first-order perturbation theory in one of the small parameters $\xi = (\alpha, \sin^2\theta_{13})$ of the SNM.  Within the solar resonance region, the simplest expressions accurate to $O(\alpha^2)$ are based on the $\sin^2\theta_{13}$-expanded eigenvalues.  Above the solar resonance region they are based on the $\alpha$-expanded eigenvalues. 

The procedure for obtaining the ${\it simplest}$ expressions for the oscillation probabilities for all transitions $\nu_a \leftrightarrow \nu_b$ is identified in Sect~\ref{ADA}.   As before, we  illustrate the procedure  for the specific case of $\nu_e \rightarrow \nu_\mu$ transitions. Expressions for our simplified oscillation probabilities are given in Appendix~\ref{summsim}.  

As emphasized in Ref.~\cite{zx}, accurate and reliable expressions for oscillation probabilities are essential for analysis and prediction at future neutrino facilities.  We calibrate these aspects of our results over all regions of energy and matter density of interest.  In Sect.~\ref{OPsrr},  we present comparisons within the solar resonance region, and in Sect.~\ref{OPasrr} we present comparisons above  the solar resonance region.

\section { Neutrino Dynamics }
\label{ND}

The dynamics of the three known neutrinos and their corresponding anti-neutrinos in matter is determined by the time-dependent Schroedinger equation,
\begin{eqnarray}
\label{tdse}
i\frac{d}{dt}|\nu(t)> &=& H_\nu |\nu(t)> ~,
\end{eqnarray}
where the neutrino Hamiltonian $H_\nu$, 
\beq
\label{fulh}
 H_\nu=H_{0v}+H_{1} ~,
\eeq
consists of a piece $H_{0v}$ describing neutrinos in the vacuum and a piece $H_{1}$ describing their interaction with matter.

Our formulation~\cite{jhk0} applies to neutrinos propagating in a uniform medium for interactions constant not only in space but also time, and
it assumes that that neutrinos and anti-neutrinos represented are the structureless elementary Dirac fields of the  Standard Neutrino Model~\cite{ISS}.  

\subsection{The Neutrino Hamiltonian in Matter}

Neutrinos are characterized by flavor $f (\nu_e, \nu_\mu,\nu_\tau)$ and mass $m [\nu_1,\nu_2,\nu_3)$.  These  states are related  by the neutrino analog of the familiar CKM matrix, 
\beq
\nu_f &=& U \nu_m ~.
\eeq  
They are produced and detected in states of a specific flavor.

The unitary matrix  $U$ is often parametrized in terms of three mixing angles $(\theta_{12},\theta_{13},\theta_{23})$ and a phase $\delta_{cp}$ characterizing $CP$ violation. 

Using the  standard abbreviations, $s_{12}\equiv\sin{\theta_{12}}$, $c_{12}\equiv\cos{\theta_{12}}$, {\it etc},
\beq
U &\equiv& \left( \begin{array}{ccc} c_{12} c_{13} & s_{12} c_{13} & s_{13} e^{-i\delta_{cp}} \\ U_{21} & U_{22} & s_{23} c_{13} \\ U_{31}  & U_{32} & c_{23} c_{13}  \end{array} \right ) ~,
\eeq 
where,
\beq
\label{Udef}
U_{21}&=& -s_{12} c_{23} - c_{12} s_{23} s_{13} e^{i\delta_{cp}} \nonumber \\
U_{22}&=& c_{12} c_{23} - s_{12} s_{23} s_{13} e^{i\delta_{cp}} \nonumber \\
U_{31}&=& s_{12} s_{23} - c_{12} c_{23} s_{13} e^{i\delta_{cp}} \nonumber \\
U_{32}&=& - c_{12} s_{23} - s_{12} c_{23} s_{13} e^{i\delta_{cp}}  ~.
\eeq
In this representation, the index  $i=1$  of  $U_{ij}$  corresponds to the electron neutrino $\nu_e$, the index 
$i=2$ to the muon neutrino $\nu_\mu$, and the index $i=3$ to the tau neutrino $\nu_\tau$.

The perturbing Hamiltonian  $H_1$ is determined by the interaction between electron neutrino flavor states and the electrons of the medium.  Expressed as matrix, 
\beq
\label{vdef}
H_{1} &=& U^{-1} \left( \begin{array}{ccc} V & 0 & 0 \\ 0 & 0  & 0 \\ 0 & 0 & 0   \end{array} \right ) U ~,
\eeq
With $V = \pm\sqrt{2} G_F n_e $ and $n_e$ the electron number density in matter. 

\subsection{The Standard Neutrino Model}

For electrically neutral matter consisting of protons, neutrons, and electrons, the electron density $n_e$ is the same as the proton density $n_p$, 
\beq
n_e &=& n_p \nonumber \\
&=& R A ~,
\eeq
where $A=N+Z$ is the average total nucleon number density of matter through which the neutrinos propagate and $R=Z/A$ is its average proton-nucleon ratio. In the earth's mantle, the dominant constituents of matter are the light elements so $R \approx 1/2$; in the surface of a neutron star $R<<1$.  Matrix elements of $H_1$ are thus
\beq
\label{mxh1h0}
<M(k)|H_{1}|M(k')> &=& U^*_{1k} V U_{1k'} ~.
\eeq

 Using the well-known expression for $V$, we find the corresponding the (dimensionless) interaction strength ${\hat A}$ of neutrinos and anti-neutrinos with matter to be,
\beq
\label{ahatrho}
\hat A &=& \pm 6.50  ~10^{-2} R ~E[{\rm GeV}] \rho[{\rm gm/cm}^3] ~,
\eeq
with E[GeV] being the neutrino beam energy  in GeV and $\rho$[gm/cm$^3$] the average total density of matter   through which the neutrino beam passes on its way to the detector in ${\rm gm/cm}^3$. For experiments close to the earth's surface, the appropriate density is the mean density of the earth's mantle,
\beq
\rho[{\rm gm/cm}^3] &=& \rho_0 \nonumber \\
&\approx& 3 ~.
\eeq

We adopt the Standard Neutrino Model~\cite{ISS}, given next, to complete our description of the  neutrino Hamiltonian.  Most of the parameters of the SNM are consistent with global fits to neutrino oscillation data with relatively good precision~\cite{hjk1,khj2}.  These include the neutrino mass differences,
\beq
m^2_2-m^2_1 &\equiv& \delta m_{21}^2 \nonumber \\
&=& 7.6\times 10^{-5} ~{\rm eV}^2
\eeq
and
\beq
m^2_3-m^2_1 &\equiv& \delta m_{31}^2 \nonumber \\
&=& 2.4\times 10^{-3} ~{\rm eV}^2 ~, 
\eeq
which corresponds to
\beq
\label{valalpha}
\alpha &\equiv& \frac{\delta m_{21}^2}{\delta m_{31}^2} \nonumber \\
&=& 3.17\times 10^{-2} ~.
\eeq

The mixing angles $\theta$ are also determined from experiment.  In the SNM, the
value of $\theta_{23}$,
\beq
\theta_{23}&=& \pi/4 ~,
\eeq
is the best-fit value from Ref.~\cite{Dav}, and $\theta_{12}$,
\beq
\theta_{12}&=& \pi/5.4 ~,
\eeq
is consistent with the recent analysis of Ref.~\cite{Gon}.  The mixing angle $\theta_{13}$ is known to be small ($\theta_{13}<0.18$ at the 95\% confidence level), but until recently its precise value has been quite  uncertain.  Results  from the Daya Bay project~\cite{DB} have measured its value more accurately,  $\sin{\theta_{13}} \approx 0.15$, which we adopt to determine our value for $\theta_{13}$,
\beq
\theta_{13}&=& 0.151  ~.
\eeq
This fixes  $R_p \equiv \sin^2\theta_{13}/\alpha \approx 0.711$. The CP violating phase is not known at all, and determining its value will one of the major interests at future neutrino facilities.

\subsection{ Our Hamiltonian Formulation} 

Our  approach  is based on the evaluation of the time-evolution operator $S(t',t)$ using the Lagrange interpolation formula~\cite{barg}. For time-independent interactions,  $S(t',t)$ is written,
\beq
\label{smtx}
S(t',t) &=& e^{-iH_\nu (t' - t)} ~.
\eeq
It depends on time  only through the time {\it difference} $t'-t$ and mat then be written in terms of the stationary state solutions $|\nu_{mi}>$ of Eq.~(\ref{tdse}), 
\beq
\label{tunif}
S(t',t) &=& \sum_i |\nu_{mi}> e^{-iE_i (t' - t ) }<\nu_{mi}| ~.
\eeq

Because the rest masses of neutrinos are considered to be tiny, for most cases of interest including our approach, the ultra-relativistic limit, $\vec |p| >> m^2$ (we take the speed of light $c=1$) is assumed. Ultra-relativistic neutrinos of energy $E$  in the laboratory frame may be expressed,  
\beq 
\label{Evac1}
E &\approx&  |\vec p| + \frac{m^2}{2E} ~,
\eeq
where $m_i$ is its mass in the vacuum. 
 
Thus, in this limit and in dimensionless variables, 
\beq
{\hat {\bar E}}_i &\to& \frac{M_i^2 - m^{2}_1}{m^{2}_3 - m^{2}_1}
\eeq
and 
\beq 
\label{h0st2}
{\hat {\bar H}}_{0v} &\to & \left( \begin{array}{ccc} 0 & 0 & 0 \\ 0 & \alpha
& 0 \\ 0 & 0 & 1 \end{array} \right ) 
\eeq
with
\beq
\label{aldef}
\alpha \equiv \frac{m_2^2-m_1^2}{m_3^2-m_1^2} ~.
\eeq
The distance $L$ from the source to the detector corresponding to $S(t',t)$ in Eq.~(\ref{smtx}) is
\beq
\label{Ldef}
L &=& t'-t ~.
\eeq

The time-evolution operator, Eq.~(\ref{smtx}), expressed in dimensionless variables is then,
\beq
\label{smtx1}
S(L) &=& e^{-i  H_\nu (t' - t)} \nonumber \\
&=& e^{2 i {\hat {\bar E}}^0_1 \Delta_L}  e^{-2 i  {\hat {\bar H}}_\nu \Delta_L} ~,
\eeq
where ${\hat {\bar H}}_\nu $ is the full neutrino Hamiltonian expressed in dimensionless variables, and where 
\beq
\label{Deldeff}
\Delta_L &\equiv& \frac{L(m^2_3-m^2_1)}{4E} ~.
\eeq
Our formulation is summarized in Appendix~\ref{SLTH}. 

Taking  the value of  $ \delta m_{21}^2$ from  the SNM, in the high-energy limit $\Delta_{L}$ [ Eq.~(\ref{Deldeff})]  becomes,  
\beq
\label{Deldefn}
\Delta_L &\approx&  3.05 \times 10^{-3} \frac { L[{\rm Km}] }{E[{\rm GeV}]} ~,
\eeq 
with  $ L[{\rm Km}] $  the baseline in $\rm {Km}$.  From Eqs.~(\ref{Deldefn},\ref{ahatrho}), we find
\beq
\Delta_L \hat A &=&  \pm 1.92  ~10^{-4} R  ~\rho[{\rm gm/cm}^3]   L[{\rm Km}]  ~,
\eeq
or
\beq
\label{DelhatA1}
 L[{\rm Km}]  &=&  \pm 5.05 ~10^{3} R    \frac{ \Delta_L \hat A} {  ~\rho[{\rm gm/cm}^3]  } ~.
\eeq

Neutrinos at rest in a neutron star were recently considered in Ref.~\cite{K}.  It was recognized that the only modification required was to take  the  non-relativistic limit of the vacuum Hamiltonian.  In our approach, this would mean taking 
\beq 
\label{Evac2}
E^0_i &\to&  m_i +  \frac{p^2}{2m_i} ~,
\eeq
rather than Eq.~(\ref{Evac1}). In particular, the eigenvalues of the interacting Hamiltonian continue to be expressed as the solution of the  same cubic equation [see Eqs.~(34,35) of Ref~\cite{jhk0}] but now with $\alpha \to {\hat \alpha}$,
\beq
\label{aldef1}
{\hat \alpha} \equiv \frac{m_2-m_1}{m_3 - m_1}  ~.
\eeq

Because it is quite straightforward to obtain the eigenvalues by solving  the cubic equation numerically,  if all one requires is an answer, the  preferable  approach would be to obtain the oscillation probability in our Hamiltonian formulation using the expressions given in Appendix~\ref{SLTH} with the coefficients given in Appendix~\ref{a:TCoefw}.  The advantage of  the analytical result we pursue in the remainder of the paper is the insight into the underlying physics that simple analytical  results provide.

\section{ Small Parameter Expansions of Observable Oscillation Probabilities }
\label{simplifyOP}

Our goal is  to obtain the simplest possible expressions for the  ${\it observable}$  neutrino oscillation probabilities accurate to about 1\%.  This is accomplished using the small-parameter expansions of  the exact analytical expressions of  our   formulation.   The small-parameter  expansion is facilitated by eliminating  $\sin^2\theta_{13}$  in favor of $\alpha$, writing $\sin^2\theta_{13} = R_p \alpha$, and then expanding  in $\alpha$.   Because $R_p \approx 0.711$ in the SNM, a Taylor expansion in $\alpha$ avoids having to expand separately in  $\alpha$ and  $\sin^2\theta_{13}$ to simplify expressions.  

\subsection{ The Observable Oscillation Probabilities  }
\label{ObOP}

The observable oscillation probabilities are expressed in terms of the partial oscillation probabilities defined in Ref.~\cite{jhk0} and summarized in Appendix~\ref{SLTH} of this paper.  Our methods are applicable to  these  oscillation probabilities for all transitions $\nu_a \to \nu_b$.  

One of the observable oscillation probabilities, $P^{+ab}(\Delta_L,\hat A) $, is the sum over three of the four partial oscillation probabilities defined in Eq.~(\ref{POPLdef1b}) and is symmetric under $a \leftrightarrow b$,
\beq
\label{pplusdef}
P^{+ab}(\Delta_L,\hat A)   &\equiv&   P^{ab}_{cos\delta} (\Delta_L,\hat A) + P^{ab}_{cos^2\delta} (\Delta_L,\hat A) \nonumber \\
&+& P^{ab}_{0} (\Delta_L,\hat A) ~.
\eeq
The structure of $P^{+ab}(\Delta_L,\hat A) $ is rather complicated, and simplifying this quantity is the focus of this paper. 

The other observable oscillation probability, $P^{ab}_{sin\delta} (\Delta_L,\hat A) $, is defined in Eq.~(\ref{a:ppsin1c2}).  It is antisymmetric under $a \leftrightarrow b$.  Because $P^{ab}_{sin\delta} (\Delta_L,\hat A) $ is both exact and already sufficiently simple, it will not be discussed further in this paper.

\bigskip

\subsection{ Leading Terms of  the $\alpha$-Expanded $ w^{(ab)}_{i,p} $ }
\label{f2ltw}

A key quantity of our   formulation~\cite{jhk0} is the set of coefficients $w^{(ab)}_{i,p}$  given explicitly in Ref.~\cite{jhk1}. These coefficients are polynomials in $\alpha$, and  for the purpose of the present work, it is rather important to note that  for many transitions $\nu_a \to \nu_b$,  the terms in the polynomial expressions for  $ w_{i;p}^{(ab)}$ with the lowest powers in $\alpha$ (the ``leading terms") cancel against each other when  ${\hat {\bar E}}_\ell \approx 1$ or ${\hat {\bar E}}_\ell \approx  0$.  One or both of these conditions hold for at least one or two of the eigenvalues over nearly the entire range of $\hat A$.  The consequence is that if we express the oscillation probabilities in terms of the original set of coefficients $w_{i;p}^{(ab)} $, three powers of  $\alpha$ must be retained to achieve $O(\alpha^2)$ accuracy, whereas we might have expected to retain only two.  Polynomials with two powers of $\alpha$ are, of course, simpler than those with three.

Polynomials with $O(\alpha^2)$ accuracy  with two powers of $\alpha$ do, in fact,  exist.  To find them, it is important to recognize  that there is nothing unique about the original set of coefficients $w^{(ab)}_{i,p} $.  We could have chosen any set of coefficients say, $w^{(ab)}_{i,p^-} $, as long as the partial oscillation probabilities remain invariant under $w^{(ab)}_{i,p} \to w^{(ab)}_{i,p^-}$.  An example of such a transformation is,
\beq 
\label{w23altdef}
w_{i;0^-}^{(ab)}    &\equiv& w_{i;0}^{(ab)} + w_{i;1}^{(ab)}  + w_{i;2}^{(ab)}  \nonumber \\
w_{i;1^-}^{(12)}    &\equiv& w_{i;1}^{(12)}  + w_{i;2}^{(12)}  \nonumber \\
w_{i;2^-}^{(12)}    &\equiv& w_{i;2}^{(12)} ~.
\eeq
Clearly, the partial oscillation probabilities are invariant under this transformation since,
\beq
\label{expw23}
{\hat {\bar w}}_i^{ab}[\ell] &\to& (w_{i;0^-}^{(ab)}  +  w_{i;1^-}^{(ab)}  ({\hat {\bar E}}_\ell - 1) + w_{i;2^-}^{(ab)} \nonumber \\
&\times&  {\hat {\bar E}}_\ell  ({\hat {\bar E}}_\ell - 1) )  \Delta {\hat {\bar E}}[\ell]  \nonumber \\
&=& (w_{i;0}^{(ab)} + w_{i;1}^{(ab)} ~{\hat {\bar E}} _{\ell} + w_{i;2}^{(ab)} ~{\hat {\bar E}} _{\ell}^2 ) \nonumber \\
&\times&  \Delta {\hat {\bar E}}[\ell] ~.
\eeq

The important points to note are, first, that the terms in the polynomial expressions for $ w_{i;p^-}^{(ab)}$, $p=0,1,2$ having the lowest powers in $\alpha$  cancel only in isolated cases.  Secondly,  the quantity  ${\hat {\bar w}}_i^{ab}[\ell]$ is expressed equivalently in terms of both $ w^{(ab)}_{i,p^-} $ and    $w^{(ab)}_{i,p} $.  Thus, the simplest expressions for the oscillation probabilities of $O(\alpha^2)$ accuracy are found from the first two powers in $\alpha$ in the expansion of $ w^{(ab)}_{i,p^-} $, whereas the same accuracy using $w^{(ab)}_{i,p}$ would require  the first three powers of $\alpha$. The coefficients  $ w^{(ab)}_{i,p^-} $ and    $w^{(ab)}_{i,p} $ are, of course, no longer equivalent when they are truncated.

Next, we compare calculations of an approximate oscillation probability  using the first two leading terms of  $ w^{(e\mu)}_{i,p^-}$  with  an exact oscillation probability calculated with $w_{i}^{ab} [\ell] $.

\begin{figure}
\centerline{\epsfig{file=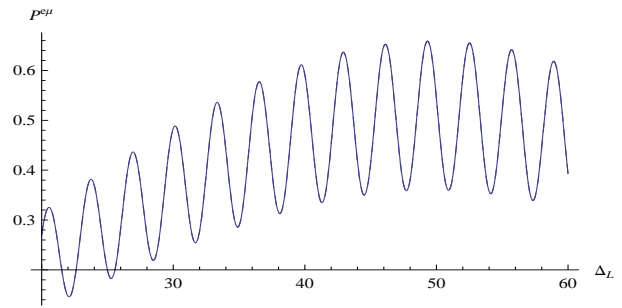,height=4cm,width=8.cm}}
\caption{ Neutrino total oscillation probability $ P^{e\mu}(\Delta_L,{\hat A})$ as a function of $\Delta_L$ for ${\hat A}=0.102$. Exact result  (solid curve); approximate result (dashed curve) calculated using exact eigenvalues and the first two leading terms of $ w^{(e\mu)}_{i,p^-}$.   All parameters except $\delta_{cp}$ are determined by the SNM. }
\label{figcptThird1}
\end{figure}

\begin{figure}
\centerline{\epsfig{file=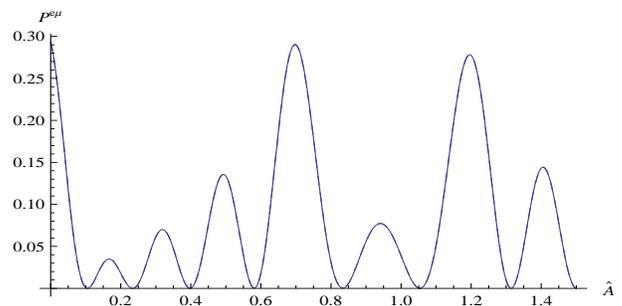,height=4cm,width=8.cm}}
\caption{ Neutrino total oscillation probability $ P^{e\mu}(\Delta_L,{\hat A})$ as a function of ${\hat A}$ for $\Delta_L=20$ and $\delta_{cp}= \pi/4$. Full theory (solid curve); approximate result (dashed curve) calculated using exact eigenvalues and the first two leading terms of $ w^{(e\mu)}_{i,p^-}$.   All parameters except $\delta_{cp}$ are determined by the SNM. }
\label{figcptThird2}
\end{figure}

Figure~\ref{figcptThird1} shows a comparison of  the two calculations of $ P^{e\mu}(\Delta_L,{\hat A})$ over the interval $20 < \Delta_L < 60$ for a value of ${\hat A} = 0.0102$.  In Fig.~\ref{figcptThird2} shows a comparison of  the two calculations   of $ P^{e\mu}(\Delta_L,{\hat A})$  over the interval $0 < {\hat A} < 1.5$ for fixed $\Delta_L = 20$.  As expected, the calculation using the first two leading terms of  $w^{(e\mu)}_{i,p^-}$ agrees very well with the exact one in both cases.

We have examined the percentage error in the partial oscillation probabilities $ P^{e\mu}_{cos}(\Delta_L,{\hat A})$ and $ P^{e\mu}_{0}(\Delta_L,{\hat A})$ calculated using the first two leading terms of  $ w^{(e\mu)}_{i,p^-}$ by comparing them numerically to the exact result.  The exact and approximate calculations agree within a fraction of a percent, except in the vicinity of zeros.  The error unavoidably diverges sufficiently close to the zeros when they are displaced by even a small amount.

We find the same cancellation just discussed for $\nu_e \to \nu_\mu$ transitions occurs for most of the other transitions 
as well.  We have tabulated 
the truncated coefficients $ w^{(ab)}_{i,p^-}$ for all these transitions in Appendix~\ref{a:TCoefw}.

 The coefficients $w^{(ab)}_{i,p^-}$, used  in conjunction with  a numerical evaluation of the neutrino mass eigenvalues, make it possible to find simple expressions for the observable oscillation probabilities  applicable to all transitions  $\nu_a \to \nu_b$ accurate to $O(\alpha^2)$ [about 1\%] throughout the entire region of interest at present and envisioned future neutrino facilities.  The mass eigenvalues are easily obtained numerically as the solution of a cubic equation.

\section{ Simple Analytic Expressions for $P^{+ab}(\Delta_L,\hat A) $ }
\label{SAX}

In this section, we develop  methods that lead to simple analytic expressions for the observable oscillation probabilities of  all transitions  $\nu_a \to \nu_b$.   The analytical results are simple, in part, because many of the terms appearing in the exact coefficients of our   approach~\cite{jhk0} are quite small  [$O(\alpha^n)$, $n \geq 3$] and, for this reason, can be eliminated.  Application of the  methods  is illustrated for flavor-changing transitions in the $(\nu_e,\nu_\mu)$ sector.

\subsection{ Simplifying $P^{+ab}(\Delta_L,\hat A) $ } 
\label{ADA}

Consider the partial oscillation probability  $P^{ab}_i(\Delta_L,\hat A) $ contributing to  $P^{+ ab}(\Delta_L,\hat A) $ in Eq.~(\ref{pplusdef}).  The idea making  it possible  to find remarkably simple expressions for  $P^{+ab}(\Delta_L,\hat A) $ can  be explained in terms of  the ${\hat {\bar w}}_i^{ab}[\ell]  $ 
corresponding to $P^{ab}_i(\Delta_L,\hat A) $ as follows.  

Expanding ${\hat {\bar w}}_i^{ab}[\ell]$ we find  
\beq
{\hat {\bar w}}_i^{ab}[\ell]  &=&   \sum_{n,i}   C_{n,i}(\ell,\alpha) {\hat A}^n ~,
\eeq
where, of course,
\beq
\label{pplusdef1}
{\hat {\bar w}}_i^{ab}[\ell]   &\equiv& (w_{i;0}^{(ab)} + w_{i;1}^{(ab)} ~{\hat {\bar E}} _{\ell} + w_{i;2}^{(ab)} ~{\hat {\bar E}} _{\ell}^2 ) \nonumber \\
&\times&  \Delta {\hat {\bar E}}[\ell] ~,
\eeq
and  $C_{n,i}(\ell,\alpha)$ is the coefficient of ${\hat A}^n$ in the expansion of  ${\hat {\bar w}}_i^{ab}[\ell]$.  Then, expanding $ C_{n,i}(\ell,\alpha)$ and retaining the first two terms in the power series in $\alpha$, 
\beq
\label{ftlt}
 C_{n,i}(\ell,\alpha) &\approx& \alpha^p(n) ( ~C(\ell)_{0;n,i} \nonumber \\
 &+& \alpha C(\ell)_{1;n,i} + ~...) ~.
\eeq
From these expressions, we can easily see  that  evaluating $P^{+ab}(\Delta_L,\hat A) $ in terms of individual $P^{ab}_i(\Delta_L,\hat A)  $ accurate to about 
1\%  fails to give the simplest expressions for $P^{+ab}(\Delta_L,\hat A)$ when $ {\hat {\bar w}}_{0}^{e \mu} [\ell] >> {\hat {\bar w}}_{cos}^{e \mu} [\ell]  $ or $ {\hat {\bar w}}_{cos}^{e \mu} [\ell] >> {\hat {\bar w}}_{0}^{e \mu} [\ell]  $.  Said otherwise, when $ {\hat {\bar w}}_{0}^{e \mu} [\ell] >> {\hat {\bar w}}_{cos}^{e \mu} [\ell]  $ or $ {\hat {\bar w}}_{cos}^{e \mu} [\ell] >> {\hat {\bar w}}_{0}^{e \mu} [\ell]  $,  simplifying the individual $P^{ab}_{i} (\Delta_L,\hat A)$ ${\it before }$ simplifying  $P^{+ab}(\Delta_L,\hat A)$ does not lead to the simplest expression for $P^{+ab}(\Delta_L,\hat A)  $.  

In this situation, the approach that does work is to  first ``consolidate" the coefficients  ${\hat {\bar w}}_{i}^{ab}[\ell]$, which means to (1)  construct the single quantities ${\hat {\bar w}}_i^{ab}[\ell] $ defined in Eq.~(\ref{hatbarwidef}); (2)  simplify ${\hat {\bar w}}_i^{ab}[\ell] $; and, (3) construct $P^{+ab}(\Delta_L,\hat A)$ from this simplified   ${\hat {\bar w}}_i^{ab}[\ell] $.  The reason why consolidation works is most easily understood by considering $C(\Delta_L,\hat A) $,
\beq
\label{sumw2}
C(\Delta_L,\hat A) &\equiv&  -   \frac{4 \Delta_L^2}{ {\hat {\bar D} } } \sum_{\ell }(-1)^\ell    {\hat {\bar w}}^{ab}[\ell]  \nonumber \\
&\times& j_0^2 (\Delta_L  \Delta {\hat {\bar E}}[\ell]) ~,
\eeq
where,
\beq
\label{hatbarwidef}
{\hat {\bar w}}^{ab}[\ell] &\equiv& \cos\delta_{cp} {\hat {\bar w}}_{cos}^{(ab)} [\ell] +\cos^2\delta_{cp} {\hat {\bar w}}_{cos^2}^{(ab)} [\ell] \nonumber \\
&+& {\hat {\bar w}}_{0}^{(ab)} [\ell]~. 
\eeq
The quantity $C(\Delta_L,\hat A)$ bears a simple relation to $P^{+ab}(\Delta_L,\hat A)$, 
\beq
 \label{sumw2a}
P^{+ab} (\Delta_L,\hat A) &=& C(\Delta_L,\hat A)~.
\eeq

The important point is that when $ {\hat {\bar w}}_{0}^{e \mu} [\ell] >> {\hat {\bar w}}_{cos}^{e \mu} [\ell]  $ or $ {\hat {\bar w}}_{cos}^{e \mu} [\ell] >> {\hat {\bar w}}_{0}^{e \mu} [\ell]  $,  the smaller coefficients are seldom among the first two leading terms of ${\hat {\bar w}}^{ab}[\ell] $ defined in Eq.~(\ref{hatbarwidef})  
and, for this reason, they  do not appear in the simplified $C(\Delta_L,\hat A) $.  However, when the individual ${\hat {\bar w}}_i^{ab}[\ell]  $ of Eq.~(\ref{pplusdef1}) are  simplified, these small terms are often  retained and appear in $P^{+ab}(\Delta_L,\hat A) $ when they are added together to obtain  $P^{+ab}(\Delta_L,\hat A) $.

The quantity ${\hat {\bar D}} $ appearing in Eq.~(\ref{sumw2}) is different for each region.  In particular,    ${\hat {\bar D}} =  \alpha C_T ( 1 - \alpha ( 1 + R_s )  )$ within the deep  solar resonance region, ${\hat {\bar D}} =  \alpha C_T (  R_s  - \alpha (1+R_s) ) / R_s^2$ within the far  solar resonance region, and $\hat {\bar D}  =   {\hat C}_\alpha ( \hat A - \alpha (c_{12}+ \hat A (c_{12} + R_p )) ) $ above the solar resonance region. 

In Eq.~(\ref{hatbarwidef}), the quantity ${\hat {\bar w}}_{i}^{(ab)} [\ell] $   is,
\beq
\label{hatbarcoef}
{\hat {\bar w}}_{i}^{(ab)} [\ell] &=& (w_{i;0^-}^{(ab)}  +  w_{i;1^-}^{(ab)}  ({\hat {\bar E}}_\ell - 1) + w_{i;2^-}^{(ab)} \nonumber \\
&\times& {\hat {\bar E}}_\ell  ({\hat {\bar E}}_\ell - 1) )\Delta {\hat {\bar E}}[\ell] ~,
\eeq
where  the  coefficients  $w_{i;p^-}^{(ab)} $ are found in Appendix~\ref{a:TCoefw}  for all transitions $\nu_a \to \nu_b$.

\subsubsection{   ${\hat C}_\alpha(\hat A) $ and $C_T(R_s)$:  Independent Variables  } 
\label{ADA}

The error in $P^{+ab} (\Delta_L,\hat A)$, so consolidated,  is most easily seen to be of order of  1\% by dropping  all terms of relative size $\alpha^n$ with $n>1$ and recognizing that $\alpha \approx 0.0317$ [see  Eq.~(\ref{valalpha})].  Said otherwise,   an  accuracy of about 1\% is achieved by retaining just  the first two terms of the $\alpha$-expansion of $ C_{n,i}(\ell,\alpha) $ ($C(\ell)_{0;n,i}$ and  $C(\ell)_{1;n,i}$).  

In this case, $P^{ab}_i(\Delta_L,\hat A) $  takes the form,
\beq
\label{pplusap}
P^{ab}_i(\Delta_L,\hat A)   &\to&  -   \frac{ 4\Delta_L^2}{ {\hat {\bar D} } }    \sum_{n,\ell} \alpha^p(n) ( C(\ell)_{0;n,i} + \alpha C(\ell)_{1;n,i} ) \nonumber \\
&\times&  {\hat A}^n j_0^2( \Delta_L \Delta {\hat {\bar E}}[\ell]  ) ~.
\eeq
Equation~\ref{pplusap}) is valid only when    ${\hat C}_\alpha $  is taken to be independent of $\hat A$, but the argument is easily generalized to take account of the dependence of  ${\hat C}_\alpha $  on  $\hat A$ (see Sect.~\ref{Dvar}).  Similar ideas apply to expressions within the solar resonance region where  ${\hat C}_\alpha \to C_T$ .

\subsubsection{ ${\hat C}_\alpha(\hat A) $ and $C_T(R_s)$:  Dependent Variables  }
\label{Dvar}

Of course, ${\hat C}_\alpha$ and $C_T$ are actually independent variables. We next consider the more realistic case  where  ${\hat C}_\alpha$ depends on $\hat A$ and $C_T$ depends on $R_s$.  

Although this discussion is somewhat technical, it explains in detail how we found the simplified expressions for $P^{+e\mu}_i(\Delta_L,\hat A) $ given in Appendix~\ref{SPEXP}, on which the numerical results of Sects.~\ref{OPsrr} and \ref{OPasrr} documenting their accuracy is based.  Because this discussion is general, it applies to all transitions $\nu_a \to \nu_b$. Although we assume that we are above the solar resonance region, the argument applies equally as well within the solar resonance region by replacing ${\hat C}_\alpha \to C_T$.

When ${\hat C}_\alpha \to {\hat C}_\alpha(\hat A) $, we first factor $C(\Delta_L,\hat A) $, Eq.~(\ref{sumw2}).  One of these  factors is proportional to $1/C^3_\alpha(\hat A) $. 
This dependence arises from the $\xi$-expanded eigenvalues appearing in Appendix~B of Ref~\cite{jhk1}  and associated with the term $w_{i;2}^{(ab)}~{\hat {\bar E}} _{\ell}^2   \Delta {\hat {\bar E}}[\ell]$.  The product  of the other factors is proportional to   $C^m_\alpha(\hat A) $, $m= 0,1,3,5, ... $.

We then replace each  term proportional to an even power, say $2n$,  of  ${\hat C}_\alpha(\hat A) $  by 
 ${\hat C}_\alpha(\hat A)^{2n}  \to ((1 - \hat A )^2 + 4 \alpha R_p \hat A)^n$.  Similarly, each  term proportional to an odd power, say $2n+1$,  of  ${\hat C}_\alpha(\hat A) $  by 
 ${\hat C}_\alpha(\hat A)^{2n+1}  \to ((1 - \hat A )^2 + 4 \alpha R_p \hat A)^n {\hat C}_\alpha(\hat A)$.

 After these replacements,  a straightforward generalization of  Eq.~(\ref{pplusap}) gives,
 \begin{widetext}
\beq
\label{pplusapX1}
C(\Delta_L,\hat A)    &\to&  -   \frac{ 4\Delta_L^2}{ {\hat {\bar D} } C^3_\alpha}    \sum_{n,\ell} \alpha^p(n) ( (C(\ell)_{0;n} + \alpha C(\ell)_{1;n} ) + {\hat C}_\alpha (C'(\ell)_{0;n} + \alpha C'(\ell)_{1;n} ) )  {\hat A}^n j_0^2( \Delta_L \Delta {\hat {\bar E}}[\ell]  ) ~,
\eeq
\end{widetext}
where $C(\ell)_{0;n},C(\ell)_{1;n}$ and $C'(\ell)_{0;n},C'(\ell)_{1;n}$ are the two sets of first two leading terms of ${\hat {\bar w}}^{ab}[\ell]$.

 \subsubsection{ Summary and Discussion} 
\label{SandD}

To summarize, by expressing  $P^{+ab}(\Delta_L,\hat A)$ in terms of  the single coefficient $ {\hat {\bar w}}^{ab}[\ell]  $ in Eq.~(\ref{hatbarcoef}), the entire set of terms $ {\hat {\bar w}}_{0}^{e \mu} [\ell]  $ and $ {\hat {\bar w}}_{cos}^{e \mu} [\ell]  $ are collected together with the consequence that
the first two leading terms of $P^{+ab}(\Delta_L,\hat A)$ is  $O(\alpha^2)$. 

At the same time, some of the higher-order terms that would have been retained by considering  $ {\hat {\bar w}}_{0}^{e \mu} [\ell]  $ and $ {\hat {\bar w}}_{cos}^{e \mu} [\ell]  $ independently are naturally discarded, leading to the ${\it simples}t$ result of  $O(\alpha^2)$.  It is the ${\it simples}t$ result of  $O(\alpha^2)$ because  the combination has fewer terms when one of the two $ {\hat {\bar w}}_{i}^{e \mu} [\ell]  $ is smaller than the other.  Inspection of the coefficients given in Appendix D indicates that this is often the case. 
  
Since the first two  leading terms constitute the simplest and most accurate expressions for the oscillation probability, this method is guaranteed to lead uniquely to  ${\it the}$ simplest and most accurate expression for $P^{+ab}(\Delta_L,\hat A)$ to about 1\%, which is what we set out to find.   We note in passing that we have managed to cast ``simplicity" into mathematical language.  Some might find this interesting, since ``simplicity is normally considered to be  a subjective concept. 

The resulting oscillation probabilities were seen to  be accurate to a few percent over all regions of interest at present and envisioned future neutrino facilities.   These analytic oscillation probabilities  are vastly more accurate than the familiar analytic expressions on the interval  $0 < \hat A < \alpha$ and for values of $\hat A$ extending from $ 0.35$ well into the asymptotic region, $\hat A >> 1$.  While somewhat  less accurate than numerical simulations,  our approximate expressions are sufficiently accuracy to  obviate the need for exact  computer simulations in many circumstances.  
The accuracy of the approximate $P^{+ab}(\Delta_L,\hat A)$  found in this way is sufficient for analysis and prediction of experiments at present and future neutrino facilities.  Furthermore, the accuracy is easy improved applying a well-defined correction procedure.

Note the following caveats. Depending on the values  of the parameters of the SNM, for specific regions it may happen that (1) some pieces of the first two leading terms are small enough to drop to maintain a specific accuracy goal;  (2) the terms of relative order $\alpha^2$ may happen to be anomaly large and thus retained to maintain an accuracy goal.

These  results may be used to define ``effective" partial oscillation probabilities.  The  ``effective" partial oscillation probability  $P_{cos\delta}^{ab}(\Delta_L,\hat A)$ would represent  the dependence of  $P^{+ab}(\Delta_L,\hat A)$ on $\delta_{cp}$.  Similarly,  an  ``effective" partial oscillation probability  $P_{cos^2\delta}^{ab}(\Delta_L,\hat A)$ would represent  the dependence of  $P^{+ab}(\Delta_L,\hat A)$ on $\delta^2_{cp}$, and   an  ``effective" $P_{0}^{ab}(\Delta_L,\hat A) $ would represent terms independent of $\delta_{cp}$.  These effective partial oscillation determine, in turn,  effective ${\hat {\bar w}}_{i}^{e \mu} [\ell]$. Note, however, that these effective partial oscillation probabilities are not guaranteed to be the simplest to $O(\alpha^2)$.

\section{ Simplified Partial Oscillation Probabilities,   $\hat A < 0.1$}
\label{OPsrr}

In this section, we consider the simplified oscillation probability over the interval $0 < \hat A <  \hat A_f$, where,  in this paper,  $\hat A_f =  0.1$.   This interval  encompasses the solar resonance, which is located at $\hat A = \alpha$.  For this reason, the interval $0 < \hat A <  0.1$ is referred to in this paper as the solar resonance region.  Note that this definition of the solar resonance region differs from that adopted in Ref.~\cite{jhk0}, where $\hat A_f = 0.2$. 
 
Here, we split up the solar resonance region into two  sub-regions. The sub-region $0 < \hat A <  \alpha $ is referred to as the deep solar  region, and  the sub-region $\alpha  < \hat A <  0.1 $ as the far solar  region.  For these sub-regions, we find  it sufficient  to retain only the first  leading term of eigenvalue difference, ${\it i.e.}$,   $\Delta {\hat {\bar E}}[\ell] = \Delta {\hat {\bar E}}_0 [\ell] $, where $ \Delta {\hat {\bar E}}_0 [\ell] $ is the first term of the Taylor series expansion of  $\Delta {\hat {\bar E}}[\ell] $ in $\sin^2\theta_{13}$ appearing in Appendix~\ref{s13SRR}.

The simplified  $P^{+ ab}(\Delta_L,\hat A) $ within  the solar resonance region is  given explicitly in Appendix~\ref{summsim} by applying the procedure of Sect.~\ref{SAX}.  We examine this $P^{+ab}(\Delta_L,\hat A) $ numerically in the present section.  By comparing $P^{+ab}(\Delta_L,\hat A) $ to its exact counterpart,  we will establish that $P^{+ab}(\Delta_L,\hat A) $ describes the $\hat A$-  and $\Delta_L$-dependence of the exact oscillation probability within the solar resonance region much more reliably than those presently found in the literature.

\subsection{General  Considerations}
\label{CCC}

From the simplified expressions for $\ P^{+ab}(\Delta_L,\hat A)$ given in Appendix~\ref{summsim}, where our results  are expressed  our in terms of effective partial oscillation probabilities and effective ${\hat {\bar w}}_{i}^{e \mu} [\ell]$, we will see  that the  effective ${\hat {\bar w}}_{cos\delta}^{e \mu} [\ell]$ is generally non-vanishing. For this reason,  $P^{+ab}(\Delta_L,\hat A) $ is sensitive to the CP violating phase,  $\delta_{cp}$ throughout  this region.  We will also see there that the situation is different above the solar resonance region.

Within the solar resonance region, the expansion in  $\sin^2\theta_{13}$  is the  appropriate one to use to obtain  the eigenvalues.  We use the representation of the $\sin^2\theta_{13}$-expanded eigenvalues given in Appendix of Ref.~\cite{jhk1}.  This representation motivates splitting  the solar resonance region into the deep solar resonance region, ${\hat A}<\alpha$ and the far solar resonance region, ${\hat A}>\alpha$.

The advantage of this representation is that by making the replacement,
\beq
\label{hatAL}
\hat A &\to &   \alpha R_s \nonumber \\
{\hat C}_\theta(\hat A)  &\to & \alpha C_T(R_s) ~,
\eeq 
in the eigenvalues of  the deep solar resonance region, and, 
\beq
\label{hatAG}
\hat A &\to &   R_s /\alpha \nonumber \\
{\hat C}_\theta(\hat A)  &\to & \hat A C_T(R_s)~ ,
\eeq 
in  those of the far solar resonance region, the ${\it same}$ function of $R_s$,
\beq
C_T(R_s) &\equiv& \sqrt{1 + R_s^2 - 2 R_s \cos 2\theta_{12} } ~,
\eeq
appears in both sets of eigenvalues.

Expressions for the effective partial oscillation probabilities given in Appendix~\ref{summsim} have been simplified using the properties that for all $\hat A > 0$, $R_s $ is positive and less than or equal to one and that 
\beq 
C_T(R_s) &=& 1+ O(\alpha) ~.
\eeq
These properties have  definite advantages for simplifying the analytic expressions we present below.  In these expressions, we sometimes use $C_T$ as an abbreviation for   $C_T(R_s)$.

\subsubsection{ The Bessel functions  $j_0^2( \Delta_L \Delta {\hat {\bar E}}[\ell]  ) $ and $\hat { \bar D} $  }

Within the solar resonance region, we find it sufficient to evaluate the Bessel functions and $\hat { \bar D} $ in terms of the first term of the Taylor series expansion of the eigenvalue difference   $\Delta {\hat {\bar E}}[\ell]$. The first term $\Delta {\hat {\bar E}}_0 [\ell] $   is given in Appendix~\ref{s13SRR}.

Accordingly,  the Bessel functions are expressed as,
\beq
j_0^2( \Delta_L \Delta {\hat {\bar E}}[\ell]  ) &=&  \frac{ \sin^2 \Delta_L   \Delta {\hat {\bar E}}_0 [\ell]  }{  \Delta {\hat {\bar E}}^2_0 [\ell]  }~,
\eeq
within the deep solar resonance region.  The energy denominator$ \hat {\bar D}$ is expressed as, 
\beq
\label{ddsr}
 \hat {\bar D}  &=& \alpha C_T ( 1 - \alpha ( 1 + R_s )  ) ~,
\eeq
within the deep solar resonance region and,  
\beq
\label{dfsr}
 \hat {\bar D}  &=& \frac{ \alpha C_T }{ R_s^2} (  R_s  - \alpha (1+R_s) ) ~,
\eeq
within the far solar resonance region.

 Within the solar resonance region, the oscillation probability  $P^{+e\mu}(\Delta_L,\hat A) $ is
\beq
P^{+e\mu}(\Delta_L,\hat A)   &\to&  - \frac{ 4}{ {\hat {\bar D} } }  \sum_{i,\ell} (-1)^\ell  {\hat {\bar w}}_{i}^{e \mu} [\ell] \nonumber \\
&\times& \frac{ \sin^2 \Delta_L   \Delta {\hat {\bar E}}_0 [\ell]  }{  \Delta {\hat {\bar E}}^2_0 [\ell]  }~.
\eeq

Numerical results for the oscillation probability $P^{+e\mu}(\Delta_L,\hat A) $ calculated within deep the solar resonance region is give below in Sects.~\ref{s:dsr} and within the far solar resonance region in Sect~\ref{fsr}.

\subsection{ Oscillation Probability $P^{+e\mu}(\Delta_L,\hat A) $ within the Deep Solar Resonance Region }

Within the deep solar resonance region, the relationship between the variables $\hat A $ and $R_s$ is  $\hat A = \alpha  R_s$.  The energy denominator $ \hat {\bar D}$ in this region is given in Eq.~(\ref{ddsr}).  The effective ${\hat {\bar w}}_{i}^{e \mu} [\ell]$, in addition to the eigenvalue differences from which the Bessel functions and energy denominator  are calculated, are given  in Table~\ref{hatwDeepSolar} of Appendix~\ref{summsim}.

\subsubsection{ Numerical Results in the Deep Solar Resonance Region}
\label{s:dsr}

\begin{figure}[h!]
\centerline{\epsfig{file=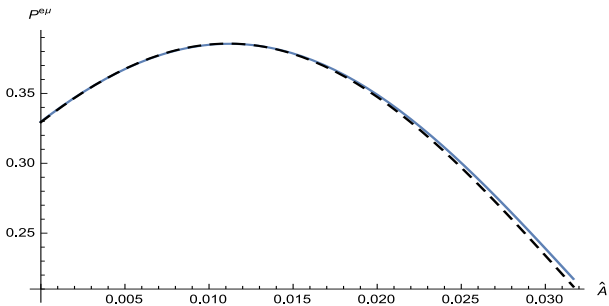,height=4cm,width=8.cm}}
\caption{$P^{+e\mu}(\Delta_L,\hat A)   $  for $\Delta_L= 60 $ over the interval $0  < \hat A < \alpha$ for neutrinos in matter.  All parameters except $\delta_{cp}$ are determined by the SNM.  Exact result (solid curve); simplified result within deep solar resonance region calculated as described above (medium-dashed curve). }
\label{figcptThird3}
\end{figure}

\begin{figure}
\centerline{\epsfig{file=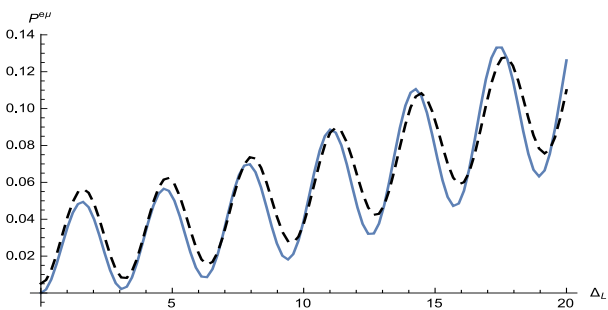,height=4cm,width=8.cm}}
\caption{$P^{+e\mu}(\Delta_L,\hat A)   $  for $\hat A = \alpha/2$  over the interval $0  < \Delta_L < 60$ for neutrinos in matter.  All parameters except 
$\delta_{cp}$ are determined by the SNM.  Exact result (solid curve); simplified result within deep solar resonance region calculated as described above (medium-dashed curve). }
\label{figcptThird4}
\end{figure}

Figures~\ref{figcptThird3} and \ref{figcptThird4} show  $P^{+e\mu}(\Delta_L,\hat A) $  compared to the exact oscillation probability as a function of $\hat A$ and $\Delta_L$, respectively.   We see from these figures that  $P^{+e\mu}(\Delta_L,\hat A) $ and the exact oscillation probability agree to a high level of accuracy within the deep solar resonance region.  

The other point of interest here is the extent to which $P^{+e\mu}(\Delta_L,\hat A) $ is an improvement over the familiar approximate formulations.  This  may be assessed  by comparing the $\hat A$ and $\Delta_L$ dependences of the present theory and the full oscillation probability of AHLO.

This assessment  of their $\hat A$ dependence follows from a comparison of Fig.~\ref{figcptThird3} to Fig.~3 of Ref.~\cite{jhk1}, which shows  the AHLO and exact oscillation probabilities as a function of $\hat A$. We see from Fig.~3 of Ref.~\cite{jhk1} that the full oscillation probability of AHLO overestimates the exact oscillation probability by about a factor of two over the interval $0  < \hat A < \alpha$, indicating that our simplified  $P^{+e\mu}(\Delta_L,\hat A) $ is a significant improvement over that of AHLO  over this interval.

An assessment of  the $\Delta_L$ dependence requires a comparison of the  $\Delta_L$ dependence of the AHLO and exact oscillation probabilities.  We have made such a comparison  within the deep solar resonance region for $\hat A \approx \alpha/3$, finding hat the AHLO result agrees with the exact result comparatively well out to $\Delta_L \approx 30$, where they begin to diverge.  The divergence continues to grow as $\Delta_L$  increases.

We see from this result and Fig.~\ref{figcptThird3} that our simplified  $P^{+e\mu}(\Delta_L,\hat A) $ agrees quite well with  AHLO  for small values of  $\Delta_L$ but not for the larger values, where all three Bessel functions interfere.

We thus find  that   $P^{+e\mu}(\Delta_L,\hat A) $ is more accurate than the full expression of AHLO by examining the $\hat A$ and the $\Delta_L$ dependence of the oscillation probability.  We conclude that   $P^{+e\mu}(\Delta_L,\hat A) $ is a significant improvement over the full expression of AHLO.

\subsection{ Oscillation Probability  within the Far Solar Resonance Region}
\label{fsr}

Within the far solar resonance region,  the relationship between the variables $\hat A $ and $R_s$ is  $\hat A = \alpha / R_s$.  The energy denominator in this region is   $ \hat {\bar D}$ given in Eq.~(\ref{dfsr}).  The effective ${\hat {\bar w}}_{i}^{e \mu} [\ell]$, in addition to the eigenvalue differences from which the Bessel functions  and energy denominator are calculated are calculated are  given  in Table~\ref{hatwFarSolar} of Appendix~\ref{summsim}.

\subsubsection{ Numerical Results in the Far Solar Resonance Region}
\label{dsr}

\begin{figure}
\centerline{\epsfig{file=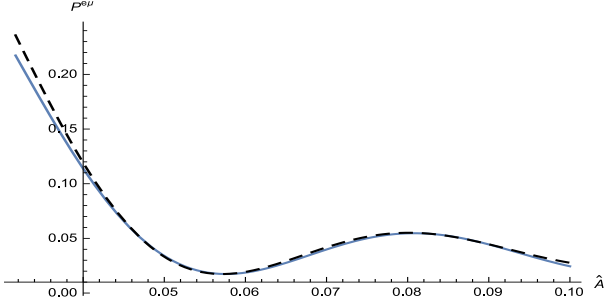,height=4cm,width=8.cm}}
\caption{$P^{+e\mu}(\Delta_L,\hat A)   $  for $\Delta_L= 60 $ over the interval $\alpha  < \hat A < 0.1$ for neutrinos in matter.  All parameters except $\delta_{cp}$ are determined by the SNM.  Exact result (solid curve); simplified result within Far solar resonance region calculated as described above (medium-dashed curve). }
\label{figcptThird5}
\end{figure}

\begin{figure}
\centerline{\epsfig{file=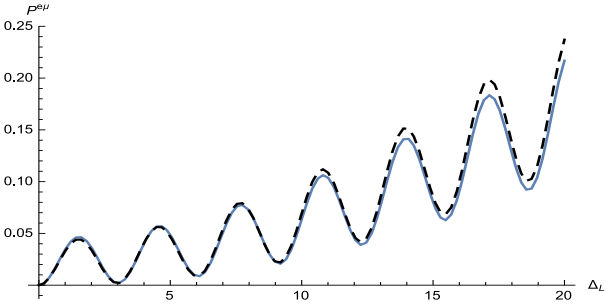,height=4cm,width=8.cm}}
\caption{$P^{+e\mu}(\Delta_L,\hat A)   $  for $\hat A = \alpha/2$  over the interval $0  < \Delta_L < 20$ for neutrinos in matter.  All parameters except 
$\delta_{cp}$ are determined by the SNM.  Exact result (solid curve); simplified result within far solar resonance region calculated as described above (medium-dashed curve). }
\label{figcptThird6}
\end{figure}

Figures~\ref{figcptThird5} and \ref{figcptThird6} show  $P^{+e\mu}(\Delta_L,\hat A) $  compared to the exact oscillation probability as a function of $\hat A$ and $\Delta_L$, respectively.   We see from these figures that  $P^{+e\mu}(\Delta_L,\hat A) $ and the exact oscillation probability agree to a high level of accuracy within the far solar resonance region.

As before, the other point of interest is again the extent to which $P^{+e\mu}(\Delta_L,\hat A) $ is an improvement over the familiar approximate formulations.  This  may be assessed  by comparing the $\hat A$ and $\Delta_L$ dependences of the present theory and the full oscillation probability of AHLO.

This assessment  of their $\hat A$ dependence follows from a comparison of Fig.~\ref{figcptThird5} to Fig.~4 of Ref.~\cite{jhk1}, which shows  the AHLO and exact oscillation probabilities as a function of $\hat A$.  Comparing Fig.~\ref{figcptThird5} to Fig.~4 of Ref.~\cite{jhk1}, one sees  that  the full result of AHLO  does reasonably well for $\alpha < \hat A < 0.1$, but our simplified  result [medium-dashed curve in Fig.~\ref{figcptThird5}] does even better.  This indicates that 
whereas our simplified  $P^{+e\mu}(\Delta_L,\hat A) $ might have some advantage  over AHLO, this advantage is marginal.

An assessment of  the $\Delta_L$ dependence requires a comparison of the  $\Delta_L$ dependence of the AHLO and exact oscillation probabilities.  We have made such a comparison  within the far solar resonance region for $\hat A \approx 0.05$ over the interval $5 < \Delta_L < 55$ and finding that the AHLO result agrees with the exact result comparatively well over this interval, even for the larger values, where all three Bessel functions interfere.

Although our simplified $P^{+e\mu}(\Delta_L,\hat A) $ is just marginally more accurate than the full expression of AHLO  the full result of Ref.~\cite{ahlo} is exceedingly complex~\cite{jhk1}.  Thus, within the far solar resonance region, our simplified $P^{+e\mu}(\Delta_L,\hat A) $ is preferable to the full result of Ref.~\cite{ahlo} because it is much simpler and more easily calculated than the full result of AHLO.

\section{ Simplified Oscillation Probabilities, $\hat A > 0.1$}
\label{OPasrr}

In this section, we consider the simplified $P^{+ab}(\Delta_L,\hat A) $ for values of  $\hat A$ above the solar resonance region. As defined in this paper, the solar resonance region encompasses the interval $0 < \hat A <  \hat A_f$, where $\hat A_f =  0.1$.  Note that this definition of the solar resonance region differs from that defined  in Refs.~\cite{jhk0,jhk1}, where $\hat A_f = 0.2$. 
 
We split up the interval  $\hat A > 0.1$ into four sub-regions.  The sub-region $0.1 < \hat A < 0.35$ is referred to as the lower transition region, and $0.35 < \hat A < \hat A_2$ as  the upper transition region. The upper and lower transition sub-regions together constitute  the transition region as defined in  Refs.~\cite{jhk0,jhk1}.  The sub-region  $\hat A_2 < \hat A < 1.2$ is referred to as the atmospheric resonance region, and the sub-region   $\hat A > 1.2$ as the asymptotic region. The  atmospheric resonance and asymptotic sub-regions are both defined here as in Refs.~\cite{jhk0,jhk1}. For these sub-regions, we find it necessary to retain both the first and the second leading terms of the eigenvalue differences, ${\it i.e.}$, $\Delta {\hat {\bar E}}[\ell]   = \Delta {\hat {\bar E}}_0 [\ell] ( 1 +r[\ell] ) $, where $ \Delta {\hat {\bar E}}_0 [\ell] $ and $r[\ell] $ are given in Appendix~\ref{s13ASRR}.

The simplified  $P^{+ ab}(\Delta_L,\hat A) $ is  given explicitly in Appendix~\ref{summsim} by applying the procedure of Sect.~\ref{SAX}.  We examine this $P^{+ab}(\Delta_L,\hat A) $ numerically in the present section. By comparing $P^{+ab}(\Delta_L,\hat A) $ to its exact counterpart,  we will establish that $P^{+ab}(\Delta_L,\hat A) $ describes the $\hat A$-  and $\Delta_L$-dependence of the exact oscillation probability within the solar resonance region much more reliably than those presently found in the literature.

\subsection{General  Considerations}
\label{CC}

We find that in all sub-regions except for the lower transition sub-region, the effective  $P_{cos}^{+e\mu}(\Delta_L,\hat A) $ vanishes.  Consequently, the entire sensitivity to the CP-violating phase $\delta_{cp}$ for these values of $\hat A$ arises from the partial oscillation probability $P_{sin\delta}^{ab}(\Delta_L,\hat A) $ given in Eq.~(\ref{a:ppsin1c2}).  We have made no attempt to simplify $P_{sin\delta}^{ab}(\Delta_L,\hat A) $ because the exact expression for $P_{sin\delta}^{ab}(\Delta_L,\hat A) $ is already quite simple in form.  Expressions for $ P^{+ab}(\Delta_L,\hat A)$ are found in Appendix~\ref{summsim}.

Above  the solar resonance region, we use  the $\alpha$-expanded representation of  ${\hat {\bar E}}_\ell$ to obtain our simplified $ P^{+ab}(\Delta_L,\hat A)$.  This $\alpha$-expanded representation appears both in Ref.~\cite{jhk0} and in an appendix of Ref.~\cite{jhk1}. 

\subsubsection{ The Bessel functions  $j_0^2( \Delta_L \Delta {\hat {\bar E}}[\ell]  ) $ and $\hat { \bar D} $  }

Above the solar resonance resonance, we found it necessary to retain both the first and second terms of the Taylor series expansion of the eigenvalue difference, ${\it i.e.}$,   $\Delta {\hat {\bar E}}[\ell] = \Delta {\hat {\bar E}}_0 [\ell] ( 1 +r[\ell] ) $, where $\Delta {\hat {\bar E}}_0 [\ell] $ is the first term of the expansion of  $\Delta {\hat {\bar E}}[\ell] $ in $\alpha $ appearing in Appendix~\ref{s13ASRR}, and $\Delta {\hat {\bar E}}_0 [\ell] r[\ell] ) $ is the second.

Accordingly, the Bessel functions are expressed,
\beq
\label{Bessasrr}
j_0^2( \Delta_L \Delta {\hat {\bar E}}[\ell]  ) &=&  \frac{ \sin^2 \Delta_L   \Delta {\hat {\bar E}}_0 [\ell] (1+r[\ell]) }{  \Delta {\hat {\bar E}}^2_0 [\ell]  (1+r[\ell]) }~.
\eeq
Equation~(\ref{Bessasrr}) may be expanded in $r[\ell]$  using Eq.~(\ref{Besssim}).  Equation~(\ref{Besssim})  is an excellent approximation for values of $\Delta_L$ appropriate above the solar resonance region, and some may find the expanded form more convenient. 
 All results shown in this section use the expanded form. 
 
Because we find it necessary to retain both the first and the second leading terms of the eigenvalue differences, ${\it i.e.}$, $\Delta {\hat {\bar E}}[\ell]   = \Delta {\hat {\bar E}}_0 [\ell] ( 1 +r[\ell] ) $, the energy denominator   $\hat {\bar D}$ becomes,
\beq
\label{asrr}
 \hat {\bar D}  &=&  {\hat C}_\alpha ( \hat A - \alpha (c_{12}+ \hat A (c_{12} + R_p )) ) ~.
\eeq

\subsection{ Lower  Transition  Region, $0.1 < \hat A < 0.35 $ }
\label{lowerTR}

For the lower transition region, $0.1 < \hat A < 0.35$, we find it necessary to make a correction $\delta P^{+e\mu}(\Delta_L,\hat A)   $ that depends on both an effective  $\delta {\hat {\bar w}}_{0}^{e \mu} [\ell]$ and an effective  $\delta {\hat {\bar w}}_{cos}^{e \mu} [\ell]$.  This correction  must be added to  ${\hat {\bar w}}_{cos}^{e \mu} [\ell]$, which is defined in Table~\ref{hatwAboveSolar}. in order to accurately describe the dependence of  $P^{+e\mu}(\Delta_L,\hat A)   $ on $\Delta_L$. Consequently,  the corrected $P^{(+)e\mu}(\Delta_L,\hat A) $  becomes a bit more complicated below $\hat A = 0.35$ and acquires a sensitivity to $\delta_{cp}$.

By including  $\delta  P^{+ab}(\Delta_L,\hat A)$, the  accuracy of he corrected $P^{+ab}(\Delta_L,\hat A)$  given improves also for some distance into the far solar resonance region.  For this reason, the solar resonance region has been redefined in this paper to cover the interval $0 < \hat A < 0.1$.

\subsubsection{ Numerical Results}

Numerical results for the  simplified $P^{+ab}(\Delta_L,\hat A) + \delta P^{+ab}(\Delta_L,\hat A) $ within the lower transition region are  presented in this section. The quantity $P^{+ab}(\Delta_L,\hat A) $ is given  in Eq.~(\ref{ppluswasrr}).  It is evaluated in terms of $ {\hat {\bar w}}_{0}^{e \mu} [1] $ from Table~\ref{hatwAboveSolar} with $D {\hat {\bar w}}_{0}^{e \mu} [1]  \to 0 $.

\begin{figure}
\centerline{\epsfig{file=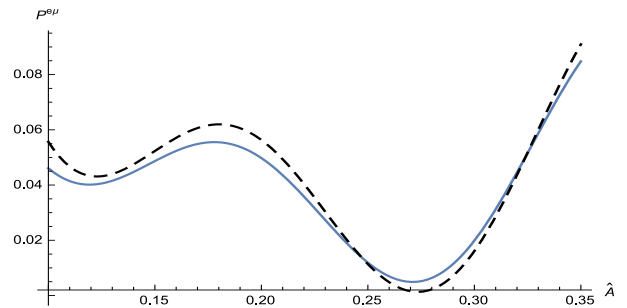,height=4cm,width=8.cm}}
\caption{ $P^{+e\mu}(\Delta_L,\hat A)   $ for $\Delta_L = 17 $ over the interval $0.1 < \hat A < 0.35 $  for neutrinos in matter.   All parameters 
 are determined by the SNM with $\delta_{cp} = \pi/4$.   Exact result (solid curve); simplified result within lower transition region calculated as described above (medium-dashed curve). }
\label{figcptThird7}
\end{figure}

\begin{figure}
\centerline{\epsfig{file=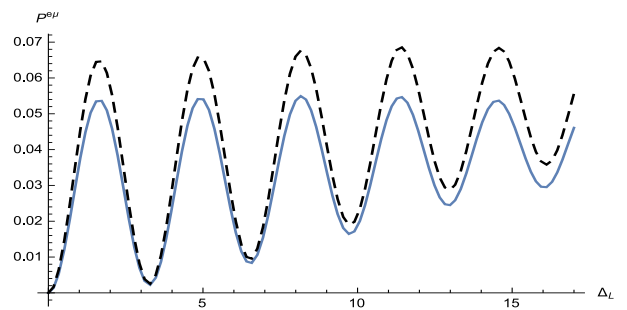,height=4cm,width=8.cm}}
\caption{ $P^{+e\mu}(\Delta_L,\hat A)   $ for $\hat A = 0.1$ over the interval $0 < \Delta_L < 17$  for neutrinos in matter.   All parameters 
 are determined by the SNM with $\delta_{cp} = \pi/4$.  Exact result (solid curve); simplified result within lower transition region calculated as described above (medium-dashed curve). }
 \label{figcptThird8}
\end{figure}

Figure~\ref{figcptThird7} shows the dependence of the simplified $P^{+ab}(\Delta_L,\hat A) + \delta P^{+ab}(\Delta_L,\hat A) $  on $\hat A$ and 
Fig.~\ref{figcptThird8} its  dependence  on $\Delta_L$.  In both figures, the simplified results are compared to the exact result~\cite{jhk1}.

We see from Fig.~\ref{figcptThird7}  that  the $\hat A$ dependence of the simplified  $P^{+e\mu}(\Delta_L,\hat A) $ and the exact oscillation probability agree moderately well.

Comparing our simplified expression for $P^{+ab}(\Delta_L,\hat A) + \delta P^{+ab}(\Delta_L,\hat A) $  shown in Fig.~\ref{figcptThird7} of this paper to Figs.~4 and 5 of of Ref.~\cite{jhk1}, one sees that the results given here are superior to Freund's patched result and comparable or superior to  his full result over the interval $0.1 < \hat A < 0.35 $.  Our results are definitely superior to the full results of AHLO~\cite{ahlo} shown in Figs.~16 and 17 of Ref.~\cite{jhk1}.

\subsection{ Upper Transition  Region, $0.35 < \hat A < \hat A_2 $ }
\label{upperTR}

We next give results we find for the effective oscillation probabilities within the  region $0.35 < \hat A < \hat A_2 $.  
The effective  ${\hat {\bar w}}_{0}^{e \mu} [\ell] $ is given in Table~\ref{hatwAboveSolar}.

\subsubsection{ Numerical Results}

\begin{figure}
\centerline{\epsfig{file=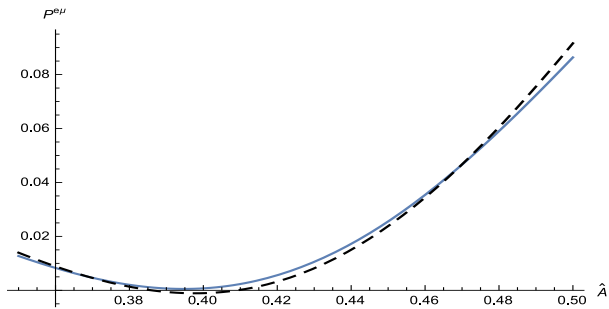,height=4cm,width=8.cm}}
\caption{ $P^{+e\mu}(\Delta_L,\hat A)   $ for $\Delta_L = 10 $ over the interval $0.35 < \hat A < \hat A_2 $  for neutrinos in matter.   All parameters 
 are determined by the SNM.   Exact result (solid curve); simplified result within upper transition region calculated as described above (medium-dashed curve). }
\label{figcptThird9}
\end{figure}

Numerical results for the  simplified $P^{+ab}(\Delta_L,\hat A) $ presented in this section are evaluations of Eq.~(\ref{pplusdef}) with   $ {\hat {\bar w}}_{0}^{e \mu} [1] $ given in Table~\ref{hatwAboveSolar} and,  of course,  $D {\hat {\bar w}}_{0}^{e \mu} [1]  \to 0 $.  Figure~\ref{figcptThird9} shows the dependence of the simplified $P^{+ab}(\Delta_L,\hat A)$ on $\hat A$ and Fig.~\ref{figcptThird10} its  dependence  on $\Delta_L$.  In both figures, the simplified results are compared to the exact result~\cite{jhk1}. These figures show that both the $\hat A$ dependence and the $\Delta_L$ dependence of the simplified results compare favorably to the exact results.

Comparing Fig.~\ref{figcptThird9} to results shown in Fig.~5 of Ref.~\cite{jhk1}, one sees that  our simplified results are superior to Freund's patched result throughout the upper transition region  and comparable to Freund's full result~\cite{jhk1}.   Our results are definitely superior to the full results of AHLO~\cite{ahlo}.

\subsection{ Atmospheric Resonance  Region: $\hat A_2 < \hat A <  1.2$ }
\label{atmR}

Following Sect.~\ref{upperTR},  we find  that  within the atmospheric resonance region $P^{+e\mu}(\Delta_L,\hat A) $ has the form given in 
Eq.~(\ref{ppluswasrr}).   The term   $D {\hat {\bar w}}_{0}^{e \mu} [1]$ is  given in Eq.~(\ref{a:D1} and needed for accuracy within this region, and ${\hat {\bar w}}_{0}^{e \mu} [1] $ is given in Table~\ref{hatwAboveSolar}. In obtaining this result we found  that both of exceptional cases (1) and (2)  discussed in Sect.~\ref{SandD} apply to this region.

\subsubsection{ Numerical Results}

\begin{figure}
\centerline{\epsfig{file=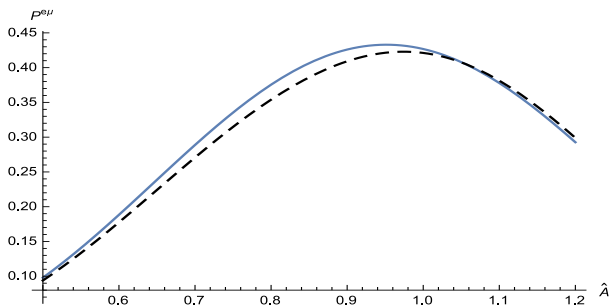,height=4cm,width=8.cm}}
\caption{ $P^{+e\mu}(\Delta_L,\hat A)   $ for $\Delta_L = 4$ over the interval $\hat A_2 < \hat A < 1.2 $  for neutrinos in matter.   All parameters 
are determined by the SNM.   Exact result (solid curve); simplified result within atmospheric resonance region  calculated as described above (medium-dashed curve). }
\label{figcptThird10}
\end{figure}

\begin{figure}
\centerline{\epsfig{file=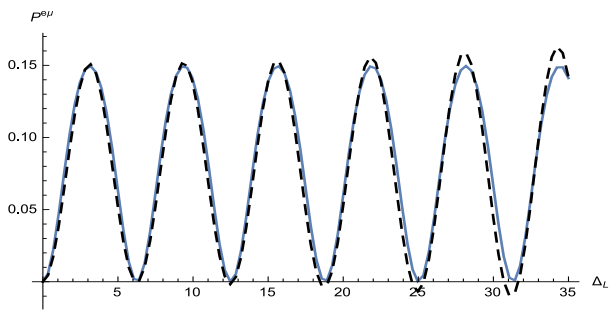,height=4cm,width=8.cm}}
\caption{ $P^{+e\mu}(\Delta_L,\hat A)   $  for $\hat A =0.8$ over the interval $0 < \Delta_L < 35$  for neutrinos in matter.   All parameters 
 are determined by the SNM.  Exact result (solid curve); simplified result within upper transition region calculated as described above (medium-dashed curve). }
\label{figcptThird11}
\end{figure}

Figures~\ref{figcptThird10} and \ref{figcptThird11} show numerical results for the first two leading terms of $P^{+e\mu}(\Delta_L,\hat A) $ in the region of the atmospheric resonance.   We see here that the dependence of  $P^{+e\mu}(\Delta_L,\hat A)   $ on both $\Delta_L$ and $\hat A$ agree well with the exact oscillation probability within this region.

Figure~6 of Ref.~\cite{jhk1} compares the $\hat A$ dependence  of the full oscillation probability of Freund to that of the oscillation probability of our   formulation evaluated with $\alpha$-expanded eigenvalues.  From this comparison, we see  that  the full oscillation probability of Freund is discontinuous at the position of the atmospheric resonance. Putting this together with Fig.~\ref{figcptThird10} leads to the conclusion that the first two leading terms of  our simplified oscillation probability is in better agreement with the exact  oscillation probability than the full oscillation probability of Freund.

Additionally, as suggested by Fig.~\ref{figcptThird11}, the $\Delta_L$ dependence of  our simplified oscillation probability agrees well with that of the exact  oscillation probability both above and below  the atmospheric resonance,

Additionally, we have found that below the atmospheric resonance the $\Delta_L$ dependence of Freund's full result~\cite{f} agrees well with the $\Delta_L$ dependence of the oscillation probability of our   formulation with $\alpha$-expanded eigenvalues, with  it's magnitude agreeing with that of the oscillation probability of our   formulation with $\alpha$-expanded eigenvalues.

Above  the atmospheric resonance, the $\Delta_L$ dependence of  our simplified oscillation probability agrees well with the $\Delta_L$ dependence of the oscillation probability of our   formulation with $\alpha$-expanded eigenvalues, but it's magnitude exceeds that of the oscillation probability of our   formulation with $\alpha$-expanded eigenvalues.

Figure~6 of Ref.~\cite{jhk1} shows that within the atmospheric resonance region the patched oscillation probability and the  full AHLO oscillation probability  are significantly larger than the exact oscillation probability and, for this reason, we conclude that Freund's patched oscillation probability and the  full AHLO oscillation probability need not be seriously considered further in the atmospheric resonance region.

\subsection{ Asymptotic  Region: $\hat A > 1.2$ }
\label{asymR}

We next give results we find for the effective oscillation probabilities within the  region $\hat A > 1.2$. The procedure given in Sect.~\ref{SAX} applied within the asymptotic region once again leads to an expression for  $P^{+e\mu}(\Delta_L,\hat A) $ having the same form that it has   in the upper  transition region,  Eq.~(\ref{ppluswasrr}).  The effective  ${\hat {\bar w}}_{i}^{e \mu} [\ell] $ is given in Table~\ref{hatwAboveSolar}  and,  of course,  $D {\hat {\bar w}}_{0}^{e \mu} [1]  \to 0 $.

\subsubsection{ Numerical Results}

\begin{figure}[h!]
\centerline{\epsfig{file=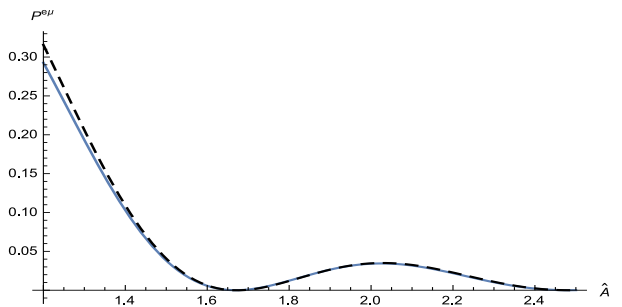,height=4cm,width=8.cm}}
\caption{ $P^{+e\mu}(\Delta_L,\hat A)   $  for $\Delta_L=4$ over the interval $1.2 < \hat A < 2.5$ for neutrinos in matter.   All parameters 
are determined by the SNM.   Exact result (solid curve); simplified result within asymptotic region calculated as described above (medium-dashed curve). }
\label{figcptThird12}
\end{figure}

\begin{figure}[h!]
\centerline{\epsfig{file=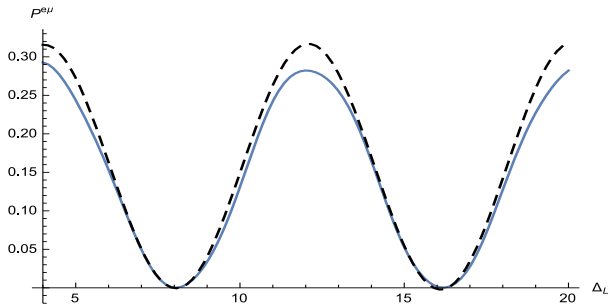,height=4cm,width=8.cm}}
\caption{ $P^{+e\mu}(\Delta_L,\hat A)   $  for $\hat A = 1.2 $ over the interval $4  < \Delta_L < 20$ for neutrinos in matter.   All parameters 
are determined by the SNM.   Exact result (solid curve); simplified result within asymptotic region calculated as described above (medium-dashed curve). }
\label{figcptThird13}
\end{figure}

Figures~\ref{figcptThird12} and \ref{figcptThird13} show  $P^{+e\mu}(\Delta_L,\hat A) $  compared to the exact oscillation probability as a function of $\hat A$ and $\Delta_L$, respectively, within the asymptotic region.   We see from these figures that  $P^{+e\mu}(\Delta_L,\hat A) $ and the exact oscillation probability agree to a high level of accuracy within this region.

As before, the other point of interest is again the extent to which $P^{+e\mu}(\Delta_L,\hat A) $ is an improvement over the familiar approximate formulations.  This  may be assessed  by comparing the $\hat A$ and $\Delta_L$ dependences of the present theory and the full oscillation probability of Freund.

This assessment  of their $\hat A$ dependence follows from a comparison of Fig.~\ref{figcptThird12} to Fig.~7 of Ref.~\cite{jhk1}, which shows  Freund's full  oscillation probability and the exact oscillation probability as a function of $\hat A$.  Comparing Fig.~\ref{figcptThird12} to Fig.~7 of Ref.~\cite{jhk1}, one sees  that  Freund's full  oscillation probability is a very good description of the exact result in the asymptotic region.  This indicates that our simplified  $P^{+e\mu}(\Delta_L,\hat A) $ offers very little advantage  over  Freund's full  oscillation probability..

An assessment of  the $\Delta_L$ dependence requires a comparison of the  $\Delta_L$ dependence of  Freund's full  oscillation probability and the exact oscillation probability.  We have made such a comparison for $\hat A = 1.2$ over the interval $ 1.2 < \Delta_L < 20$ finding that Freund's full oscillation probability agrees very well with the exact oscillation probability.

The $\Delta_L$ dependence our simplified $P^{+e\mu}(\Delta_L,\hat A) $ agrees less  well with the exact  oscillation probability within the asymptotic region by about 10\%  than does Freund's full result near the peaks of $P^{+e\mu}(\Delta_L,\hat A) $.  Otherwise, the two agree comparably as well. Whether or not one adopts our simplified $P^{+e\mu}(\Delta_L,\hat A) $ is therefore a matter of whether this 10\% difference is significant for the intended purpose.

Figure~7 of Ref.~\cite{jhk1} shows that within the asymptotic region the patched oscillation probability of Freund is significantly larger than the exact oscillation probability and, for this reason, we conclude that Freund's patched oscillation probability need not be seriously considered further in the asymptotic region.

\subsection{ Discussion }

We have found the  remarkable result in this section that ${\hat {\bar w}}_{0}^{e \mu} [1]$ vanishes in  all regions above the lower transition region. Accordingly, we  establish an important result, namely  that  $P^{+ab}(\Delta_L,\hat A)$ is almost completely independent of the CP violating phase,  $\delta_{cp}$, for $\hat A > 0.35$.

Another important result is that the expressions we find for $P^{(+)e\mu}(\Delta_L,\hat A) $ are  very accurate above the solar resonance region. These expressions are identical for the first two leading of ${\hat {\bar w}}_{0}^{e \mu} [1]  $ in all regions except for the lower transition region and the vicinity of the  atmospheric resonance. These expressions are also very simple  for $\hat A > 0.35$.  They are given in Table~\ref{hatwAboveSolar}. Above  the lower transition region, we find, of course, that the effective $ P_{cos}^{ab}(\Delta_L,\hat A) $ vanishes.

Within the lower transition region, the correction term contains both effective ${\hat {\bar w}}_{0}^{e \mu} [\ell] $ and effective ${\hat {\bar w}}_{cos}^{e \mu} [\ell] $ leading to more complicated expressions for  $P^{+ab}(\Delta_L,\hat A)  $ and showing that it  begins to acquire a sensitivity to $\delta_{cp}$ within this region.

It should be emphasized that only in a few instances are these effective partial oscillation probabilities suitable approximations to the exact ones, even though the  symmetric oscillation probability $P^{+ab}(\Delta_L,\hat A) $ obtained by summing the three effective ones as in Eq.~(\ref{pplusdef}) does  satisfy our accuracy goal. This discussion highlights the fact that even though determination of simple and accurate expressions is purely algebraic, finding the {\it simplest} expressions is hardly a trivial process.

 \section{Summary and Conclusions}
 
 In this paper, we have determined simple and accurate algebraic expressions for the observable oscillation probabilities characterizing three coupled Dirac neutrinos described by the SNM and convenient for quantitative predictions and analyses of high quality data becoming available at new neutrino facilities.   The expressions we found are based on our recently published Hamiltonian formulation. 
 
 The procedure leading to our  simple and accurate expressions for the oscillation probabilities relies  on  on two main ideas.
The first  is that many terms constituting the coefficients $w^{(ab)}_{i,p}$ of our original Hamiltonian formulation~\cite{jhk1} are small enough to be  neglected because of subtile cancellations that occur among these coefficients.  We determined a transformation for all transitions $\nu_a \to \nu_b$ that led to a new set of coefficients $w^{(ab)}_{i,p^-}$.  These are  essential for obtaining simple results for the oscillation probability accurate to $O(\alpha^2)$ because they are  devoid of such cancellations.   The second main idea forms the basis of an efficient and general procedure leading to the ${\it simplest}$ expressions for oscillation probabilities accurate to $O(\alpha^2)$.

The methods established here are general.  Simple and accurate expressions for the neutrino oscillation probabilities for all $\nu_a \to \nu_b$ transitions are easily obtained with the information given in the appendix of this paper.

We  illustrated this procedure and calibrated its   for the specific case of $\nu_e \leftrightarrow \nu_\mu$ transitions.   These transitions correspond to the experimental situation for which data exists and are the only transitions for which analytic results have been attempted in the literature.

Figures presented in the present  paper established the    accuracy and reliability of our simple  algebraic expressions for the observable oscillation probabilities.    Our results are vastly more accurate  than  results that have been available in the literature for some time   throughout the entire region of interest at present and envisioned future neutrino facilities.  This accuracy is     adequate for analysis and prediction of experiments being considered at future neutrino facilities. Our  expressions are only slightly more complicated  than the familiar ones.

Our result relies on neutrino mass eigenvalues  evaluated using first-order perturbation theory in one of the small parameters $\xi = (\alpha, \sin^2\theta_{13})$ of the SNM.  Within the solar resonance region, the simplest expressions accurate to $O(\alpha^2)$ are based on the $\sin^2\theta_{13}$-expanded eigenvalues,  and above the solar resonance region they are based on the $\alpha$-expanded eigenvalues.

Alternatively, even more accurate results follow using  this  information in conjunction with  a numerical evaluation of the neutrino mass eigenvalues, which are easily obtained as the solution of a cubic equation.  Because this method lacks a fully analytic representation, it is less useful than the analytic result, since, as in the case for all numerical solutions, it is more difficult to obtain insight into the physics in this way.

\begin{widetext}
\appendix

%tohere2

\section{Our   Formulation}
\label{SLTH}

Our   formulation~\cite{jhk0} gives explicit, analytic expressions for the partial oscillation probabilities in terms of the mixing angles, ${\hat A}$, and the neutrino eigenvalues, ${\hat{\bar E}}_\ell$.

\subsection{Full Oscillation Probability in Our   Formulation}

 One of the partial oscillation probabilities,  $P_{\sin\delta}^{ab} $, is antisymmetric under $a\leftrightarrow b$,
\beq
\label{a:ppsin1c2}
P_{sin\delta}^{ab}(\Delta_L,\hat A)  &=&  \sin {\delta_{cp} }    \Delta_L^3  \alpha (1-\alpha)  (-1)^{a+b}  \cos{\theta_{13}} \sin{ 2 \theta_{12}} \sin{2\theta_{13}} \sin{2\theta_{23}}  \Pi_{\ell}  j_0 ( \Delta_L \Delta {\hat{\bar E}}[\ell] )  ~.
\eeq
The other three are  individually symmetric,
\beq
\label{POPLdef1b}
P_{cos\delta}^{ab}(\Delta_L,\hat A)   &=& -  \cos \delta_{cp}  \frac{4 \Delta_L^2}{ {\hat {\bar D} } }    \sum_\ell (-1)^\ell  {\hat {\bar w}}^{ab}_{cos}[\ell] 
 j_0^2 (  \Delta_L \Delta {\hat{\bar E}}[\ell] )  \nonumber \\
P_{cos^2\delta}^{ab}(\Delta_L,\hat A)   &=& -  \cos^2 { \delta_{cp} }  \frac{4 \Delta_L^2}{ {\hat {\bar D} } }    \sum_\ell (-1)^\ell  {\hat {\bar w}}^{ab}_{cos^2}[\ell]   j_0^2 (  \Delta_L \Delta {\hat{\bar E}}[\ell] ) \nonumber \\
 \nonumber \\
P_{0}^{ab}(\Delta_L,\hat A)   &=& -   \frac{ 4\Delta_L^2}{ {\hat {\bar D} } }    \sum_\ell (-1)^\ell {\hat {\bar w}}^{ab}_0[\ell]   j_0^2 (  \Delta_L \Delta {\hat{\bar E}}[\ell] )   ~,
\eeq
where the sum runs over $\ell=1~,2~,3$,
\beq
\label{hatdeldef}
{\hat {\bar D} } &=&  \Delta \hat {\bar E}[1] \Delta \hat {\bar E}[2] \Delta \hat {\bar E}[3] ~,
\eeq
and, the quantity $\Delta_L $ appearing in Eqs.~(\ref{a:ppsin1c2},\ref{POPLdef1b}) is defined in Eq.~(\ref{Deldeff}).
In Eq.(\ref{POPLdef1b}),
 \beq
\label{hatbarw}
{\hat {\bar w}}^{ab}_{i}[\ell]   &=&  (w_{i;0}^{(ab)} + w_{i;1}^{(ab)} ~{\hat {\bar E}} _{\ell} + w_{i;2}^{(ab)} ~{\hat {\bar E}} _{\ell}^2 )  \Delta {\hat {\bar E}}[\ell]  ~.
\eeq

The bracket notation is the same as that used throughout   our work~\cite{jhk0}.

 The total oscillation probability for neutrinos is then,
\beq
\label{Pmemu0}
\mathcal{P}(\nu_b  \rightarrow \nu_a) &=& \delta(a,b) +  P_0^{ab} + P_{\sin\delta}^{ab} + P_{\cos\delta}^{ab} \nonumber \\
&+& P_{\cos^2\delta}^{ab} ~.
\eeq

Because  the masses are ordered so that $ m_3 > m_2 > m_1$ and eigenvalues do not cross, it follows that $ {\hat {\bar E}}_3 > {\hat {\bar E}}_2 >{\hat {\bar E}}_1$ for all $|{\hat A}|$.  Consequently,  $\Delta {\hat {\bar E}[\ell]}$ as well as  ${\hat {\bar D} }$ are all positive. 

The partial oscillation probabilities  in Eqs.~(\ref{a:ppsin1c2},\ref{POPLdef1b}) have been expressed  as explicit functions of  $\Delta_L$ and $\hat A$ to indicate that the entire dependence on the neutrino beam energy $E$, the baseline $L$, and the medium properties occurs   through  $\Delta_L$ and ${\hat A}$.
This follows  from their definitions   in Eqs.~(\ref{ahatrho},\ref{Deldeff}), respectively.

The matrices  $ w_{i;n}^{ab} $ appearing in Eq.~(\ref{POPLdef1b})  take the form~\cite{jhk0}.
\beq
\label{a:Pcos2g}
&& w_{cos^2;n}  =   \sin^2{2\theta_{23}} \left( \begin{array}{ccc} 0 & 0 & 0 \\ 0  & w^{(22)}_{cos^2;n}& - w^{(22)}_{cos^2;n} \\ 0 & - w^{(22)}_{cos^2;n}& w^{(22)}_{cos^2;n} \end{array} \right ) ~,
\eeq 
\beq
\label{wcoseva0l} 
&& w_{cos;n}  = \sin 2\theta_{23}  \left( \begin{array}{ccc} 0  & w^{(12)}_{cos;n} & - w^{(12)}_{cos;n}  \\ w^{(12)}_{cos;n} & w^{(22)}_{cos;n}   & w^{(23)}_{cos;n} \\ - w^{(12)}_{cos;n} & w^{(23)}_{cos;n}   & w^{(33)}_{cos;n}     \end{array} \right ) ~,
\eeq 
and
\beq
\label{w0eva0l}
&& w_{0;n} =  \left( \begin{array}{ccc} w_{0;n}^{(11)}   & w_{0,n}^{(12)}  & w_{0;n}^{(13)} \\ w_{0;n}^{(12)} & w_{0;n}^{(22)}   & w_{0;n}^{(23)}  \\ w_{0;n}^{(13)} & w_{0;n}^{(23)} & w_{0;n}^{(33)} \end{array} \right ) ~.
\eeq
The partial oscillation probabilities given in Eqs.~(\ref{a:ppsin1c2},\ref{POPLdef1b}) are ${\it exact}$ when evaluated in terms of the exact eigenvalues [see Eq.~(40) of Ref~\cite{jhk0}] and the full expressions for the coefficients  ${ w}_{i;n}^{(ab)} $ [see  Appendix~C  of Ref~\cite{jhk1}].

The usefulness of the oscillation probabilities  follows from various  interconnections among them. 
One of these is that the exchange of initial and final states in the oscillation probability or neutrinos (antineutrinos) is equivalent to letting $\delta_{cp}\rightarrow -\delta_{cp}$. Thus, the result for the inverse reaction $\mathcal{P}(\nu_b  \rightarrow \nu_a)$ is found by exchanging $(a,b)$ in Eq.~(\ref{Pmemu0}).
 Since  $ P_{\sin\delta}^{ab} $ is antisymmetric under the exchange of $(a,b)$, and $ P_0^{ab}$, $P_{\cos\delta}^{ab}$ and 
$P_{\cos^2\delta}^{ab}$ symmetric, it follows that $P^{ba}$ is given by
\beq
\label{Pmmume0}
\mathcal{P}(\nu_b  \rightarrow \nu_a) &=& \delta(a,b) +  P_0^{ab} - P_{\sin\delta}^{ab} + P_{\cos\delta}^{ab} \nonumber \\
&+& P_{\cos^2\delta}^{ab} ~.
\eeq

In analogy to Eq.~(\ref{Pmemu0}), we may express the oscillation probability for antineutrinos as
\beq
\label{Pbarmemu0}
&&\mathcal{P}({\bar \nu}_a  \rightarrow {\bar \nu}_b) \equiv  {\bar P}^{ab} (\Delta_L,\hat A)  \nonumber \\
&=& \delta(a,b) + { \bar P}_0^{ab} + {\bar P}_{\sin\delta}^{ab} + {\bar P}_{\cos\delta}^{ab} + {\bar P}_{\cos^2\delta}^{ab}
  ~,
\eeq
where the bared probabilities for anti neutrinos are obtained from the unbarred for neutrinos by replacing $\delta_{cp}\rightarrow -\delta_{cp}$ and ${\hat A} \rightarrow - {\hat A}$. Because the energies of antineutrinos are different from those of the neutrinos in matter, we can expect $P^{ab} \neq {\bar P}^{ab}$ in this situation.

Again applying the rule that exchange of initial and final states is accomplished by letting $\delta_{cp}\rightarrow -\delta_{cp}$, the oscillation probability $\mathcal{P}({\bar \nu}_b  \rightarrow {\bar \nu}_a)$ is expressed in terms of the same four quantities,
\beq
\label{Pbarmmue0}
&&\mathcal{P}({\bar \nu}_b  \rightarrow {\bar \nu}_a) \nonumber \\
&=& \delta(a,b) +  {\bar P}_0^ {ab} -  {\bar P}_{\sin\delta}^{ab}   + {\bar P}_{\cos\delta}^{ab}  + {\bar P}_{\cos^2\delta}^{ab}
 ~.
\eeq

\section{Approximate Oscillation Probability in Our   Formulation}

As discussed in Sect.~\ref{simplifyOP}, many terms constituting the exact expressions for coefficients $w^{(ab)}_{i,p}$, which are   given explicitly in Appendix C of Ref.~\cite{jhk1}, are small and may be neglected without compromising the overall reliability of our result.  However, it is not straightforward to recognize the cancelling terms because of subtile cancellations that occur among these coefficients.  The source of these cancellations is identified in Sect.~\ref{f2ltw} above and, based on this understanding, we determine a transformation  leading to a new set of coefficients $w^{(ab)}_{i,p^-}$,

\beq
\label{expw23a}
{\hat {\bar w}}_{i}^{ab} [\ell] &=& (w_{i;0^-}^{(ab)}  +  w_{i;1^-}^{(ab)}  ({\hat {\bar E}}_\ell - 1) + w_{i;2^-}^{(ab)}  {\hat {\bar E}}_\ell  ({\hat {\bar E}}_\ell - 1) )\Delta {\hat {\bar E}}[\ell] ~,
\eeq
which are devoid of cancellations. The first two leading terms of the coefficients $w^{(ab)}_{i,p^-}$   are essential for obtaining simple results for the oscillation probability accurate to $O(\alpha^2)$.  These coefficients are tabulated below for all transitions  $\nu_a \to \nu_b$.
The expressions for the approximate oscillation probability are those given in Eqs.~(\ref{a:ppsin1c2},\ref{POPLdef1b}), but  with $ w_{i;p^-}^{ab} [\ell] $ evaluated using  the first two leading terms of the coefficients ${\hat {\bar w}}_{i;p^-}^{(ab)} $.  The overall factor $K^i_{ab}$ of each coefficient ${\hat {\bar w}}_{i;p^-}^{(ab)} $   appears in Table~\ref{coef110b}.  The actual coefficient $w^{(ab)}_{i;,p^-}$ are found from those given below by multiplying them by   $K^i_{ab}$,
\beq
w^{(ab)}_{i;p^-} &\to & K^{i}_{2}  w_{i;p^-}^{(ab)} ~.
\eeq

\section{First Two Leading Terms of the Coefficients  $ w_{i;p^-}^{(ab)} $ }
\label{a:TCoefw}

The first two leading terms of the coefficients $ w^{(ab)}_{cos^2,p^-} $ are given in Sect.~\ref{a:wcos2}; those for $ w^{(ab)}_{cos,p^-} $ in Sect.~\ref{a:wcos}; and, those for $ w^{(ab)}_{0,p^-} $ in Sect.~\ref{a:w0}.

\subsection{ $ w_{p}^{(22)} (i = cos^2) $ }
\label{a:wcos2}

\beq
w_{cos^2,0^-}^{(22)} &=&  (1- \alpha +\hat A (1 - \alpha (s_{12}^2 +R_p))) \nonumber \\
w_{cos^2,1^-}^{(22)} &=&  (1-\alpha)   \nonumber \\
w_{cos^2,2^-}^{(22)} &=&   1 ~.
\eeq
The remaining coefficients $ w^{(ab)}_{cos^2,p^-} $ are found from these  as follows,
\beq
w^{(23)}_{cos^2, p^-} &=& w^{(32)}_{cos^2, p^-}  = -  w^{(22)}_{cos^2, p^-}  \nonumber \\
w^{(33)}_{cos^2, p^-} &=& w^{(22)}_{cos^2, p^-}   \nonumber \\ 
w^{(11)}_{cos^2, p^-} &=& w^{(12)}_{cos^2, p^-} = w^{(13)}_{cos^2, p^-} = w^{(21)}_{cos^2, p^-} \nonumber \\
&=& w^{(31)}_{cos^2, p^-}  = 0 ~,
\eeq

\subsection{ $ w_{p~,}^{(12)}$ $ w_{p~,}^{(23)}$ and $ w_{p}^{(22)} (i = cos) $ }
\label{a:wcos}

\beq
w_{cos;0^-}^{(12)} &=&   \alpha (c_{12}^2 -s_{12}^2)  (1-\alpha (1+R_p)) - \hat A (1-\alpha (1+R_p)  \nonumber \\     
w_{cos;1^-}^{(12)} &=&  (1-  \alpha  (1+2  s_{12}^2 +R_p) ) -   \hat A w_{cos;2^-}^{(12)}
 \nonumber \\  
w_{cos;2^-}^{(12)} &=&( 1 - \alpha( s_{12}^2+ R_p))  ~;
\eeq
\beq
w_{cos;0^-}^{(23)} &=&  c_ {12}  s_{12} ( \alpha ( c_ {12}^2 -    s_ {12}^2) - \alpha^2 (1 - R_p)  \nonumber \\
&\times& (c_ {12}^2 - 
      s_ {12}^2)) (c_ {23}^2 - s_ {23}^2) -\hat A (1 - \alpha (1 + 3 R_p))   \nonumber \\     
w_{cos;1^-}^{(23)}&=& - c_ {12}  (1 - \alpha (1 + 2 c_ {12}^2 + R_p)) s_{12} (c_ {23}^2 - 
   s_ {23}^2)  \nonumber \\ 
w_{cos;2^-}^{(23)} &=& -2  c_ {12} s_{12}  (1 - \alpha (c_ {12}^2 + R_p))  (c_ {23}^2 \nonumber \\
& -& s_ {23}^2)  ~;
\eeq
and,
\beq 
w_{cos;0^-}^{(22)} &=&  \alpha c_ {23}^2 (c_ {12}^2 - s_ {12}^2) - \alpha^2 (c_ {12}^2 - 
    s_ {12}^2) (c_ {23}^2 +  R_p s_ {23}^2) \nonumber \\
& -& \hat A (c_ {23}^2 - \alpha (c_ {23}^2 + 
       R_p (1 - 3 s_ {23}^2)))  \nonumber \\
w_{cos;1^-}^{(22)} &=& -\hat A (1 - \alpha ( s_ {12}^2  +  R_p )) + 
 s_ {23}^2 + \alpha (c_ {12}^2 - s_ {12}^2 \nonumber \\
 &-& 
    s_ {23}^2 (1 + 2 c_ {12}^2 + R_p) )  \nonumber \\
w_{cos;2^-}^{(22)} &=& 2 s_ {23}^2 + \alpha (c_ {23}^2 - 2 c_ {23}^2 s_ {12}^2 -  s_ {23}^2 - 2 R_p \nonumber \\
&\times& s_ {23}^2)  ~.
\eeq

The remaining coefficients $ w^{(ab)}_{cos,p^-} $ are found from these as follows,
\beq
w^{(13)}_{cos, p^-} &=&  w^{(31)}_{cos, p^-} = - w^{(12)}_{cos, p^-} \nonumber \\
w^{(21)}_{cos, p^-} &=& w^{(12)}_{cos, p^-} ~.
\eeq
The coefficients $ w^{(33)}_{cos,p^-}    $ are obtained from $ w_{cos,p^-}^{(22)}$ by making the replacement $\sin{\theta_{23}} \leftrightarrow \cos{\theta_{23}} $ and  flipping the overall sign.

\subsection{ $ w_{p~,}^{(12)}$ $ w_{p~,}^{(23)}$ $ w_{p~,}^{(11)}$ and $ w_{p}^{(22)} (i = 0) $ }
\label{a:w0}
\beq  
w_{0;0^-}^{(12)} &=&   \alpha (4  s_ {12}^2 c_ {12}^2 (c_ {23}^2 - \alpha (c_ {23}^2 + R_p))
   - \hat A (4 c_ {12}^2 c_ {23}^2 s_ {12}^2 + 
      4 R_p^2 s_ {23}^2 - \alpha (4 c_ {12}^4 c_ {23}^2 s_ {12}^2 - 
         R_p s_ {23}^2 (4 c_ {12}^2 s_ {12}^2 - 12 R_p s_ {12}^2  - 
            4 R_p^2)  )))   \nonumber \\
w_{0;1^-}^{(12)} &=& (4 \alpha s_ {12}^2 (c_ {12}^2 c_ {23}^2 - R_p s_ {23}^2) - \alpha^2
     s_ {12}^2 (4 c_ {12}^2 c_ {23}^2 + 
     4 R_p (c_ {23}^2 - s_ {12}^2 - s_ {23}^2 - R_p s_ {23}^2)))
-\hat A  w_ {0; 2^-}^{(12)}  \nonumber \\
w_{0;2^-}^{(12)} &=&   4 R_p s_ {23}^2 + \alpha (4 c_ {12}^2 c_ {23}^2 s_ {12}^2 - 
    4 R_p (R_p + 2 s_ {12}^2) s_ {23}^2)  ~;
\eeq
\beq
w_{0;0^-}^{(23)} &=&-\alpha (   16 \alpha c_ {12}^2 c_ {23}^2 s_ {12}^2 s_ {23}^2 +  4s_ {23}^2 c_ {23}^2 \hat A (    4 R_p -   \alpha (   4 c_ {12}^2 s_ {12}^2 + 
  4 R_p (1 + 2 R_p + s_ {12}^2))   )  \nonumber \\
  &-&  \alpha^2 (16 c_ {12}^2 c_ {23}^2 s_ {12}^2 s_ {23}^2 +   2 R_p (8 c_ {12}^2 s_ {12}^2 + 8 c_ {23}^2 s_ {23}^2 - 
   16 c_ {12}^2 c_ {23}^2 s_ {12}^2 s_ {23}^2)))    \nonumber \\
w_{0;1^-}^{(23)} &=&  -16\alpha s_ {23}^2 c_ {23}^2 (c_ {12}^2 - \alpha (c_ {12}^2 +     c_ {12}^4 + R_p))  \nonumber \\
 w_{0;2^-}^{(23)} &=&   16 c_ {23}^2  s_ {23}^2 (1 - 2 \alpha (c_ {12}^2 + R_p)) ~;
\eeq

\beq 
w_{0;0^-}^{(11)} &=& -4 \alpha  s_ {12}^2 c_ {12}^2 ( 1 - \alpha (1 + 2 R_p)) \nonumber \\
w_{0;1^-}^{(11)} &=& -4 \alpha s_ {12}^2 ( c_ {12}^2 -  R_p - \alpha (c_ {12}^2 - R_p (R_p + 2 s_ {12}^2)  )  \nonumber \\
w_{0;2^-}^{(11)} &=& - 4 R_p - \alpha (4 c_ {12}^2 s_ {12}^2 - 4 R_p (R_p + 2 s_ {12}^2))   ~;
\eeq
and,
\beq 
w_{0;0^-}^{(22)} &=& \alpha (4 \alpha c_ {23}^4 c_ {12}^2 s_ {12}^2 - \alpha^2  c_ {23}^2 (4 c_ {12}^2 c_ {23}^2 s_ {12}^2 - 
 2 R_p s_ {23}^2 (1 +  (c_ {12}^2 -  s_ {12}^2)^2) )   - \hat A ( 4 R_p c_ {23}^2 s_ {23}^2 \nonumber \\
&+& \alpha (4 c_ {12}^2 c_ {23}^4 s_ {12}^2    -  4 R_p s_ {23}^2    (c_ {23}^2 (1 + s_ {12}^2) + R_p (c_ {23}^2 - s_ {23}^2) )  )))   \nonumber \\
w_{0;1^-}^{(22)} &=& \alpha (4 c_{12}^2 c_{23}^2 s_{23}^2 
-   \alpha (2 R_p s_{23}^2 (  c_{12}^2 + c_{23}^2 - s_{12}^2 - s_{23}^2 +   c_{12}^2 c_{23}^2 (   1 + 2 s_{12}^2 +   6 s_{23}^2 - (   c_{12}^2 - s_{12}^2  )   (c_{23}^2 - s_{23}^2)  )     )  \nonumber \\
&+&    4 \hat A (R_p s_{23}^2 +  \alpha (c_{12}^2 c_{23}^2 s_{12}^2 - R_p (R_p + 2 s_{12}^2) s_{23}^2))   ) \nonumber \\
w_ {0; 2^-}^{(22)} & = & - 4  s_ {23}^2 (c_ {23}^2 - \alpha (2 c_ {12}^2 c_{23}^2 + R_p (c_ {23}^2 - s_ {23}^2))) 
\eeq

The remaining coefficients $ w^{(ab)}_{0,p^-} $ are obtained from these as follows,
\beq
w^{(21)}_{0, p^-} &=& w^{(31)}_{0, p^-}  \nonumber \\
&=& w^{(12)}_{0, p^-} ~.
\eeq
The coefficients $ w^{(33)}_{0,p^-}    $ are obtained from $ w_{0,p^-}^{(22)}$ by making the replacement $\sin{\theta_{23}} \leftrightarrow \cos{\theta_{23}} $. Likewise, $ w^{(13)}_{0,p^-}    $ are obtained from $ w_{0,p^-}^{(12)}$ the same way,  making the same replacement, $\sin{\theta_{23}} \leftrightarrow \cos{\theta_{23}} $.

\begin{table}[h!]
\caption {\label{coef110b}  Coefficients $ K^i_{ab} $ of the  coefficients $w_{i;p^-}^{(ab)}$ } 
\begin{ruledtabular}
\begin{tabular}{c c c c c c c c c}
              & $i$           &(a,b)       & $ K^i_{ab} $  \\ 
& & & \\
              &$c2$           & $ (2,2)  $  &   $\alpha^3 R_p s_{12}^2 c_{12}^2$    \\
& & & \\
              &$c$             & $ (1,2) $    &    $\alpha \frac{  \sqrt{\alpha R_p} }{2}  s_{12} c_{12}$    \\
& & & \\
              & $c$            & $(2,3)$      &    $\alpha  \sqrt{\alpha R_p}   s_{12}^2 c_{12}^2$    \\
& & & \\
              & $c$            & $(2,2)$ &  $- \alpha \sqrt{\alpha R_p}  s_{12} c_{12}$    \\

& & & \\
              & $0$            & $(1.2)$   &  $ \frac{\alpha }{4}  $    \\ 
& & & \\
              & $0$            & $(2,3)$   &  $ \frac{1 }{16}  $    \\
 & & & \\
              & $0$            & $(1,1)$  &  $ \frac{\alpha }{4}  $    \\
 & & & \\
              & $0$            & $(2,2)$   &  $ \frac{1 }{4}  $    \\
\end{tabular} \\
\end{ruledtabular} 
\end{table}

\end{widetext}
\bigskip
\section{ Small Parameter  Expansions }
\label{SPEXP}

We have shown~\cite{jhk0}  that expansions of the eigenvalues in the  small parameter $\alpha$ is troublesome in the vicinity  of the solar resonance and that expansions of the eigenvalues in the  small parameter $\sin^2\theta_{13}$  is troublesome in the vicinity  of the atmospheric resonance.   These problems are avoided in this paper by expressing     the  oscillation probability      within the solar resonance region by using       the expansion of the eigenvalues in  
$\sin^2\theta_{13}$       and above the solar resonance region by using  the expansion of the eigenvalues in  $\alpha$.

\subsection{ Quantities $ \Delta {\hat {\bar E}}_0 [\ell] $ and  $r[\ell]$}

In this section, we define the  energy differences $\Delta {\hat {\bar E}} _{0}[\ell] $ and $\Delta  {\hat {\bar E}} _{1}[\ell] $, 
\beq
\label{DeltaE}
\Delta {\hat {\bar E}} _{n}[1] &=&    {\hat {\bar E}} _{n,  3} -{\hat {\bar E}} _{n,  2}  \nonumber \\
\Delta {\hat {\bar E}} _{n}[2]& = &   {\hat {\bar E}} _{n,  3} -{\hat {\bar E}} _{n,  1}  \nonumber \\
\Delta {\hat {\bar E}} _{n}[3] &=&    {\hat {\bar E}} _{n,  2} -{\hat {\bar E}} _{n,  2}  ~.
\eeq 
For $n=0$,  $ {\hat {\bar E}} _{n,  \ell} $ is  the first term of a Taylor expansion in $\alpha$ of the  $\xi$-expanded eigenvalues ${\hat {\bar E}}_\ell$, and for 
$n=1$  $ {\hat {\bar E}} _{n,  \ell} $ is  the second term of the Taylor expansion.  

We then define from these  $\Delta {\hat {\bar E}} _{0}[\ell] $ and $\Delta  {\hat {\bar E}} _{1}[\ell] $ the  quantities $\Delta {\hat {\bar E}} [\ell] $ and $r[\ell]$,
\beq 
\label{rdef0}
\Delta {\hat {\bar E}} [\ell] &\equiv& \Delta {\hat {\bar E}}_0 [ \ell] +  \Delta {\hat {\bar E}}_1 [ \ell] \nonumber \\
&=& \Delta {\hat {\bar E}}_0 [\ell] ( 1 +r[\ell] ) ~,
\eeq
where, clearly,
\beq 
\label{DErdef0}
r[\ell ] &\equiv& \Delta  {\hat {\bar E}} _1[ \ell] / \Delta {\hat {\bar E}} _0 [\ell]~.
\eeq
The quantities   $\Delta {\hat {\bar E}} _{0}[\ell] $ and $r[\ell] $ are of course different above and below the solar resonance region. 
 
Because $r[\ell]$ is small parameter, we may consider simplifying the oscillation probability by expanding pieces of  it in $r[\ell]$. The quantities we have in mind are  the Bessel function, $  j_0^2( \hat \Delta[\ell] )  $ and $\hat D$, 
\beq
\hat D &\equiv& \Delta {\hat {\bar E}}[1] \Delta {\hat {\bar E}}[2] \Delta {\hat {\bar E}}[3]  ~.
\eeq

\subsection{ Quantities $ \Delta {\hat {\bar E}}_0 [\ell] $ and  $r[\ell]$ within the Solar Resonance Region }
\label{s13SRR}

As discussed, there are advantages to splitting the solar resonance region into two sub-regions. One of these is  the deep solar resonance region,  $\hat A < \alpha$, and the other is the far solar resonance region, $\hat A > \alpha$.

\subsubsection {   $\xi = \sin^2\theta_{13}$ , $\hat A < \alpha$ }

For  $\hat A < \alpha$, we find the following expressions for  $\Delta  {\hat {\bar E}}_{0}[\ell ]$, 
\beq
\label{LeadDEs12Below}
(a) ~\Delta  {\hat {\bar E}}_{0}[1] &=& 1 - \frac{\alpha }{2} (1 + R_s + C_T ) \nonumber \\
(b) ~\Delta {\hat {\bar E}}_{0}[2] &=& 1 - \frac{\alpha }{2} (1 + R_s - C_T )  \nonumber \\
(c) ~\Delta  {\hat {\bar E}}_{0}[3] &=& \alpha C_T 
\eeq
The corresponding  parameters $r[\ell]$, defined in  Eq.~(\ref{DErdef0}), are
\beq
(a) ~r[1] &=&  \alpha^2 \frac{R_p R_s (2 + R_s)}{(1 - \alpha R_s) (2 - \alpha (1 + R_s +C_T))}  \nonumber \\
(b) ~r[2] &=&  \alpha^2 \frac{R_p R_s (4-R_s)}{(1 - \alpha R_s) (2 - \alpha (1 + R_s -C_T))}  \nonumber \\
(c) ~r[3] &=&  \alpha  \frac{R_p R_s (1 -R_s)}{C_T(1 - \alpha R_s)} ~.
\eeq

\subsubsection {  $\xi = \sin^2\theta_{13}$,  $\hat A > \alpha$ }

For $\hat A > \alpha$, we find,
\beq
\label{LeadDEs12Above}
(a) ~\Delta  {\hat {\bar E}}_{0}[1] &=& 1 - \frac{\alpha }{2 R_s} (1 + R_s + C_T ) \nonumber \\
(b) ~\Delta {\hat {\bar E}}_{0}[2] &=& 1 - \frac{\alpha }{2 R_s} (1 + R_s - C_T )   \nonumber \\
(c) ~\Delta  {\hat {\bar E}}_{0}[3] &=&  \frac{ \alpha}{R_s} C_T 
\eeq
and 
\beq
(a) ~r[1] &=& - \alpha^2 \frac{ 3 R_s R_p } { (\alpha - R_s) (2 R_s  -\alpha  (1 + R_s + C_T) )}  \nonumber \\
(b) ~r[2] &=& - \alpha^2 \frac{ 3 R_s R_p } { (\alpha - R_s) (2 R_s  -\alpha  (1 + R_s - C_T) )}  \nonumber \\
(c) ~r[3] &=& 0 ~.
\eeq

\subsection { Expressions for $ \Delta {\hat {\bar E}}_0[ \ell]$ and $r[\ell]$  Above  the Solar Resonance Region}
\label{s13ASRR}

For the $\alpha$-expanded eigenvalues, we find,
\beq
\label{LeadDEalpha}
(a) ~\Delta  {\hat {\bar E}}_{0}[1] &=&    {\hat C}_\alpha    \nonumber \\
(b) ~\Delta {\hat {\bar E}}_{0}[2] &=&   \frac{1}{2} ( 1 + {\hat A}  + {\hat C}_\alpha )  \nonumber \\ 
(c) ~\Delta  {\hat {\bar E}}_{0}[3] &=& \frac{1}{2} ( 1 + {\hat A}  - {\hat C}_\alpha ) 
\eeq
and 
\beq
\label{ralpha}
(a) ~r[1] &=& - \alpha  \frac{  s_{12}^2 }{{\hat C}_\alpha^2} (1 - \hat A)   \nonumber \\
(b) ~r[2] &=& -  \alpha  \frac{  s_{12}^2 (1-\hat A ) -{\hat C}_\alpha (1 - 3 c_{12}^2 )}{  (1 - \hat A)^2+ {\hat C}_\alpha (1 - \hat A) + 4 \alpha \hat A R_p }     \nonumber \\
(c) ~r[3] &=& -  \alpha \nonumber \\
&\times&  \frac{  s_{12}^2 (1-\hat A ) + {\hat C}_\alpha (1 - 3 c_{12}^2 )}{  (1 - \hat A)^2 -  {\hat C}_\alpha (1 - \hat A) + 4 \alpha \hat A R_p }    ~.
\eeq

\section{Observable Neutrino Oscillation Probabilities}
\label{summsim}

In Appendix~\ref{SLTH}, the full neutrino oscillation probability, $\mathcal{P}(\nu_a  \rightarrow  \nu_b) $, is expressed as the sum of two observable oscillation probabilities.   One of these, $P_{\sin\delta}^{ab} $, is anti-symmetric under the exchange $a\leftrightarrow b$ and the other, which we call  $P^{+e\mu}(\Delta_L,\hat A) $,  is symmetric. The anti-symmetric observable given in Eq.~(\ref{a:ppsin1c2}) is proportional to $\sin\theta_{cp}$, where $\delta_{cp}$ is the CP violating phase.   The exact expression for this  quantity is sufficiently simple that it requires no further simplification.  However,  the analytic expression for $P^{+e\mu}(\Delta_L,\hat A) $ is rather complicated, and its simplification is the subject of this paper. These $P^{+e\mu}(\Delta_L,\hat A) $ have been simplified by region as explained in Sect.~\ref{SAX}.  In this Appendix we summarize  our findings.

\subsection{   Regions of $\hat A$}

The oscillation probability $P^{+e\mu}(\Delta_L,\hat A) $  will be given separately for $0 < \hat A < 0.1 $, which we refer to as the solar resonance region, and the region  $\hat A > 0.1 $.  Each of these regions is subdivided into various intervals.

The oscillation probability $P^{+e\mu}(\Delta_L,\hat A) $ takes quite different forms depending upon the interval of $\hat A$ considered. The most complicated expressions for our simplified $P^{+e\mu}(\Delta_L,\hat A) $ are those describing the lower transition region.   

\bigskip

\subsection{Effective $P_{cos\delta}^{ab}(\Delta_L,\hat A)$, $P_{cos^2\delta}^{ab}(\Delta_L,\hat A)$,  and $P_{0}^{ab}(\Delta_L,\hat A)$}

As discussed  in Sect.~\ref{SandD}, ``effective" partial oscillation probabilities $P_{cos\delta}^{ab}(\Delta_L,\hat A)$, $P_{cos^2\delta}^{ab}(\Delta_L,\hat A)$,  and $P_{0}^{ab}(\Delta_L,\hat A)$ may be defined from  $P^{+e\mu}(\Delta_L,\hat A) $. The  effective partial oscillation probability  $P_{cos\delta}^{ab}(\Delta_L,\hat A)$ represents  the dependence of  $P^{+ab}(\Delta_L,\hat A)$ on $\delta_{cp}$;   the effective partial oscillation probability  $P_{cos^2\delta}^{ab}(\Delta_L,\hat A)$ represents  its  dependence on $\delta^2_{cp}$; and, the`effective $P_{0}^{ab}(\Delta_L,\hat A) $ its  terms independent of $\delta_{cp}$.

These effective partial oscillation determine, in turn,  effective ${\hat {\bar w}}_{i}^{e \mu} [\ell]$. The expressions for the effective  ${\hat {\bar w}}_{i}^{e \mu} [\ell] $ are much simpler than the full  ${\hat {\bar w}}_{i}^{e \mu} [\ell] $, as is the case for the Bessel functions.  Note, however, that neither the effective partial oscillation probabilities nor the effective ${\hat {\bar w}}_{i}^{e \mu} [\ell]$ are guaranteed to be the simplest to $O(\alpha^2)$.

As we will see, two exceptional cases,  described in Sect.~\ref{SandD}, apply to both regions of $\hat A$.  

\section{Summary of Simplified Oscillation Probabilities}
\label{summsim}

In this section, we give the simplified analytic expressions for the effective partial oscillation probabilities $P_{cos\delta}^{ab}(\Delta_L,\hat A)$, $P_{cos^2\delta}^{ab}(\Delta_L,\hat A)$,  and $P_{0}^{ab}(\Delta_L,\hat A)$ and the  effective ${\hat {\bar w}}_{i}^{e \mu} [\ell]$. Results are presented by region.

\bigskip

\begin{widetext}

\subsection{Solar Resonance Region, $0 < \hat A < 0.1 $}

The solar resonance region consists of the deep solar resonance region, $0 < \hat A < \alpha $, and the  far solar 
resonance region, $\alpha < \hat A < 0.1 $.  
Within this region, our  simplified  oscillation probability symmetric in $a\leftrightarrow b$ is of the form,

\beq
\label{ppluswssr}
P^{+ab}(\Delta_L,\hat A)  &=&  - \frac{ 4\Delta_L^2}{ {\hat {\bar D} } }  \sum_{i,\ell} (-1)^\ell   {\hat {\bar w}}_{i}^{ab} [\ell]   j_0^2 (   \Delta_L\Delta {\hat {\bar E}} [\ell] )  ~,
\eeq
where the sum runs over $i = (0,cos)$. Within the solar resonance region, we find  it sufficient  to retain only the first  leading term of eigenvalue difference, ${\it i.e.}$,   $\Delta {\hat {\bar E}}[\ell] = \Delta {\hat {\bar E}}_0 [\ell] $, where $ \Delta {\hat {\bar E}}_0 [\ell] $ is given in Appendix~\ref{s13SRR}. 

\subsubsection{ Deep Solar Resonance Region, $0 < \hat A < \alpha $}

The expressions we find for the effective  ${\hat {\bar w}}_{i}^{e \mu} [\ell] $ within the deep solar resonance region are given in Table~\ref{hatwDeepSolar}  We also give  the eigenvalue difference $\Delta {\hat {\bar E}}_0 [\ell] $.

\begin{table}[h!]
\caption {\label{hatwDeepSolar}   Deep Solar  } 
\begin{ruledtabular}
\begin{tabular}{c c c c c c c c c}
        & Quantity            &           &$\ell = 1$       & $ \ell = 2 $    &$\ell = 3$   \\ 
& & & \\
             & ${\hat {\bar w}}_{0}^{e \mu} [\ell] $         &      & $  K_0   ( 2 s_{12}^2  + R_s)  $                         &   $K_0   (  c_{12}^2  - R_s)   $     &   $0$    \\
& & & \\
          & ${\hat {\bar w}}_{cos}^{e \mu} [\ell] $         &    & $ - K_C   (  1 - \alpha (3 + R_p + 4 R_s) )  $    &    $- {\hat {\bar w}}_{cos}^{e \mu}[1]$    &   $0$    \\
& & & \\
              & $\Delta {\hat {\bar E}}_0 [\ell]  $         &  &  Eq.~(\ref{LeadDEs12Below})~(a)     &   Eq.~(\ref{LeadDEs12Below})~(b)   & Eq.~(\ref{LeadDEs12Below})~(c)  \\
\end{tabular} \\
\end{ruledtabular} 
\end{table}

\subsubsection{ Far Solar Resonance Region, $\alpha  < \hat A < 0.1$ }

Expressions for effective values of ${\hat {\bar w}}_{i}^{e \mu} [\ell] $ within the far solar resonance region are  given in Table~\ref{hatwFarSolar}. The coefficients $AA$, $K_C $, and $K_0 $ appearing in Table~\ref{hatwFarSolar} are defined as follows,
\beq
AA &\equiv& -1-3\alpha + \alpha R_p + R_s \cos 2 \theta_{12}+10 \alpha s_{12}^2  \nonumber \\
K_C &\equiv&- \cos\delta_{cp}  \frac{\alpha \sqrt{\alpha R_p}} {2 (\alpha-  R_s)^3}  s_{12} c_{12} \nonumber \\
K_0 &\equiv& \frac{\alpha^2  }{2(\alpha - R_s)^3 }  R_s ~.
\eeq

\begin{table}[h!]
\caption {\label{hatwFarSolar}  ${\hat {\bar w}}_{cos}^{e \mu} [\ell] $: Far Solar  } 
\begin{ruledtabular}
\begin{tabular}{c c c c c c c c c}
        & Quantity            &           &$\ell = 1$       & $ \ell = 2 $    &$\ell = 3$   \\ 
& & & \\
          & ${\hat {\bar w}}_{cos}^{e \mu} [\ell] /K_C$         &    & $  R_s  (R_s-2\alpha)^2 $  &   $- {\hat {\bar w}}_{cos}^{e \mu}[1]/K_C$    &  $

           \alpha^2 (  R_s C_T ( 1 - R_s \cos 2 \theta_{12} ) -  \alpha ( 3 + R_s R_p )        )

           $     \\
& & & \\
          & ${\hat {\bar w}}_{0}^{e \mu} [\ell] /K_0$         &    &$   R_s   R_p s_{23}^2 ( AA+ C_T  )          $           & $  - R_p s_{23}^2   (8 \alpha +  R_s (AA   - C_T + 2 \alpha R_p  ) 
          $  &  $  \alpha  s_{12}^2 c_{12}^2  ( R_s c^2_{23} C_T - 4  \alpha  c_{23}^2   )  $ \\
& & & \\
              &  $\Delta {\hat {\bar E}}_0[\ell]   $      &  & Eq.~(\ref {LeadDEs12Above}) ~(a)      &   Eq.~(\ref {LeadDEs12Above}) ~(b)  &     Eq.~(\ref {LeadDEs12Above}) ~(c)     \\              
\end{tabular} \\
\end{ruledtabular} 
\end{table}

\subsection{Above Solar Resonance  Region, $\hat A > 0.1$}

The regions with $\hat A > 0.1$ consist of the lower transition  region, $0.1 < \hat A < 0.35$; the upper transition region, $0.35 < \hat A <  \hat A_2$; the atmospheric resonance region, $\hat A_2  < \hat A < 1.2$; and, the asymptotic region. $\hat A  > 1.2$. Within all these regions, we find it necessary to retain both the first and the second leading terms of the eigenvalue differences, ${\it i.e.}$, $\Delta {\hat {\bar E}}[\ell]   = \Delta {\hat {\bar E}}_0 [\ell] ( 1 +r[\ell] ) $.

In all regions above the solar resonance region, except for the lower transition region, our  simplified  oscillation probability $P^{+ab}(\Delta_L,\hat A)$  has the same form with just one  component, 
\beq
\label{ppluswasrr}
P^{+ab}(\Delta_L,\hat A) &=&   \frac{ 4\Delta_L^2}{ {\hat {\bar D} } }  ( {\hat {\bar w}}_{0}^{ab} [1]  + D {\hat {\bar w}}_{0}^{e \mu} [1] ) j_0^2 (  \Delta_L\Delta {\hat {\bar E}} [1] ) ~.
\eeq
The coefficient  ${\hat {\bar w}}_{0}^{e \mu} [1]$ is given in Table~\ref{hatwAboveSolar} and $D {\hat {\bar w}}_{0}^{e \mu} [1]$,
\beq
\label{a:D1}
D {\hat {\bar w}}_{0}^{e \mu} [1] &=& \frac{ 2 \alpha^2 R_p^2 } { {\hat C}_\alpha}    s_{23}^2  ( 3   c_{12}^2   + 4  R_p )   ~,
\eeq
is a correction required  for accuracy in the atmospheric resonance region.   

The Bessel function in Eq.~(\ref{ppluswasrr})  simplified by  using  its expanded representation given in Eq.~(\ref{Besssim}) below and  taking  $\Delta {\hat {\bar E}}[1] $ from Table~\ref{hatwAboveSolar}.  We find,
\beq
\label{ppluswpp1b}
P^{+ab}(\Delta_L,\hat A) &=&   \frac{ 4\Delta_L^2}{ {\hat {\bar D} } }  ( ( {\hat {\bar w}}_{0}^{ab} [1]  + D {\hat {\bar w}}_{0}^{e \mu} [1] )  j_0^2 ( \Delta_L  {\hat C}_\alpha )    +2 r[\ell]   {\hat {\bar w}}_{0}^{ab} [1]    (j_0( 2 \Delta_L  {\hat C}_\alpha )-  j_0^2 ( \Delta_L  {\hat C}_\alpha ) ) ) ~,
\eeq
where the term $ r[\ell]  D {\hat {\bar w}}_{0}^{e \mu} [1] $ has been dropped because it is of order $\alpha^3$ and therefore quite small.  The
 short-hand notation $\tilde A \equiv 1 -  {\hat A } $ is used in Table~\ref{hatwAboveSolar} and below for the frequently occurring quantity $1 -  {\hat A }$.

\bigskip
\subsubsection{The Lower Transition Region}

Within the lower transition region, our  simplified  oscillation probability symmetric under $a \leftrightarrow b$ consists of two pieces $P^{+ab}(\Delta_L,\hat A)$ and $\delta P^{+ab}(\Delta_L,\hat A)$, where   $P^{+ab}(\Delta_L,\hat A)$ is given in Eq.~(\ref{ppluswssr}), and  $\delta P^{+ab}(\Delta_L,\hat A)$ is,

\beq
\label{ppluswpp1del}
\delta P^{+ab}(\Delta_L,\hat A) &=&  - \frac{ 4\Delta_L^2}{ {\hat {\bar D} } }  \sum_{i,\ell} (-1)^\ell  \delta {\hat {\bar w}}_{i}^{ab} [\ell]  j_0^2 ( \Delta_L\Delta {\hat {\bar E}} [\ell] ) ~,
\eeq

\begin{table}[h!]
\caption {\label{hatwlowertransition}  Effective $\delta {\hat {\bar w}}_{0}^{e \mu} [\ell] /K^0_{1,2} $, Lower Transition. $K^0_{1,2} [\ell] $ is given inTable~\ref{lowertransition}.   } 
\begin{ruledtabular}
\begin{tabular}{c c c c c c c c c}
              &    Quantity          &$\ell = 1$       & $ \ell = 2 $    &$\ell = 3$   \\   \\ 
& & & \\
              &   $\delta {\hat {\bar w}}_{0}^{e \mu} [\ell] /K^0_{1,2} $         & $ - 4   \alpha R_p s_{23}^2 \tilde A (2 (c_{12}^2 -R_p) - \tilde A (1+ s_{12}^2 + 2 R_p) )  $                &   $       0                 
                        $     &   $ 2 \alpha  \tilde A \hat A c_{12}^2 c_{23}^2s_{12}^2     $    \\
\end{tabular} \\
\end{ruledtabular} 
\end{table}

\begin{table}[h!]
\caption {\label{dhatwAboveSolar}   Effective $\delta {\hat {\bar w}}_{cos}^{e \mu} [\ell]/K^c_{1,2} $    Lower Transition   $K^c_{1,2} [\ell] $  is given inTable~\ref{lowertransition}.   } 
\begin{ruledtabular}
\begin{tabular}{c c c c c c c c c}
        &        &           &   $\delta {\hat {\bar w}}_{cos}^{e \mu} [\ell]/K^c_{1,2} $   &     &   \\ 
& & & \\
          & $\ell= 1 $          &    &$1 - \alpha (1 + 2 s_{12}^2 + R_p )  - 2  \hat A (1 - \alpha (1 + 2 s_{12}^2 + R_p ))  +  \hat A^2 (1- \alpha ( 1 + 3 R_p) )  $      &         &      \\
& & & \\
          & $\ell = 2$     &    &$ - \frac{ 1 }{ \tilde A^2 }  (1 - \alpha (1 + 2 c_{12}^2 + R_p ) +   \hat A (2 - \alpha (1 + 2 c_{12}^2 + 9  R_p ))  +  \hat A^2 (1- \alpha ( 1 + 2 c_{12}^2 + 3 R_p) )   + \alpha  R_p \hat  A^3  )                $                &       &          \\          
& & & \\
          & $\ell = 3 $        &   &$ - \frac{ 1}{ \tilde A^2 }  (1 - {\hat C}_\alpha  -   \hat A (2 - {\hat C}_\alpha  +\alpha (2 - 4 c_{12}^2 -  R_p ))  + \alpha  \hat A^2 ( 5 - 8 c_{12}^2 + 3 R_p)) +  \hat A^3 (2 - \alpha ( 4 s_{12}^2 + 9 R_p) )    
 \hat A^4 (1 - \alpha ( 1 +  R_p )  )     $                &        &            \\       
                          \\              
\end{tabular} \\
\end{ruledtabular} 
\end{table}

\begin{table}[h!]
\caption {\label{lowertransition}  Definitions of  $K^0_{1,2} [\ell] $  (Table~\ref{hatwlowertransition}) and $K^c_{1,2} [\ell] $ (Table~\ref{dhatwAboveSolar})  } 
\begin{ruledtabular}
\begin{tabular}{c c c c c c c c c}
              &  Quantity          &$\ell = 1$       & $ \ell = 2 $    &$\ell = 3$   \\          
& & & \\
              &$K^c_{1,2} [\ell] $          & $\alpha c_{12} s_{12} \sqrt {\alpha R_p }/2   $             &   $\alpha c_{12} s_{12} \sqrt {\alpha R_p }/2   $     &    $\alpha c_{12} s_{12} \sqrt {\alpha R_p }/2   $     \\
& & & \\
              &$K^0_{1,2} [\ell] $          & $\alpha R_p s_{23}^2 / \tilde A  $             &   $\alpha R_p s_{23}^2 / \tilde A  $     &    $\alpha R_p s_{23}^2 / \tilde A  $     \\                                         
\end{tabular} \\
\end{ruledtabular} 
\end{table}

\begin{table}[h!]
\caption {\label{hatwAboveSolar} ${\hat {\bar w}}_{0}^{e \mu} [\ell] $  and $\Delta {\hat {\bar E}}[\ell]  = \Delta {\hat {\bar E}}_0 [\ell] ( 1 +r[\ell] ) $ Above  Solar  } 
\begin{ruledtabular}
\begin{tabular}{c c c c c c c c c}
        & Quantity            &           &$\ell = 1$       & $ \ell = 2 $    &$\ell = 3$   \\ 
& & & \\
          & ${\hat {\bar w}}_{0}^{e \mu} [\ell] $         &    &$  -  \frac{ \alpha R_p  }{{\hat C}_\alpha}   s_{23}^2   (  \hat A  \tilde A^2   + \alpha (4 R_p  -   \tilde A ( s_{12}^2   + 8 R_p ) ) ) +D {\hat {\bar w}}_{0}^{e \mu} [1]                  $                &   $   0                         
                        $      &    $0 $        \\          
& & & \\
              &$\Delta {\hat {\bar E}}_0[\ell] $          & Eq~(\ref{LeadDEalpha})~(a)                &   Eq~(\ref{LeadDEalpha})~(b)       &   Eq~(\ref{LeadDEalpha})~(c)      \\
& & & \\
              &$r[\ell] $          & Eq~(\ref{ralpha})~(a)               &  Eq~(\ref{ralpha})~(b)     &   Eq~(\ref{ralpha})~(b)     \\          
\end{tabular} \\
\end{ruledtabular} 
\end{table}

with $\delta {\hat {\bar w}}_{0}^{e \mu} [\ell] $  given in Table~\ref{hatwlowertransition} and  $\delta {\hat {\bar w}}_{cos}^{e \mu} [\ell] $  in Table~\ref{dhatwAboveSolar}.  In Eq.~(\ref{ppluswpp1del}),  the sum over $i$ runs over the two values  $ (0,cos)$. Note that the effective $\delta {\hat {\bar w}}_{0}^{e \mu} [1] =0$. 
\bigskip

\subsubsection{Above the Lower Transition Region}

In all regions above the  lower transition region, our  simplified  oscillation probability $P^{+ab}(\Delta_L,\hat A)$  has just one  component, 
\beq
\label{ppluswasrr}
P^{+ab}(\Delta_L,\hat A) &=&   \frac{ 4\Delta_L^2}{ {\hat {\bar D} } }  ( {\hat {\bar w}}_{0}^{ab} [1]  + D {\hat {\bar w}}_{0}^{e \mu} [1] ) j_0^2 (  \Delta_L\Delta {\hat {\bar E}} [1] ) ~,
\eeq

where the coefficient  ${\hat {\bar w}}_{0}^{e \mu} [1]$ is given in Table~\ref{hatwAboveSolar} and  the quantity $D {\hat {\bar w}}_{0}^{e \mu} [1]$,
\beq
\label{a:D1}
D {\hat {\bar w}}_{0}^{e \mu} [1] &=& \frac{ 2 \alpha^2 R_p^2 } { {\hat C}_\alpha}    s_{23}^2  ( 3   c_{12}^2   + 4  R_p )   ~,
\eeq
is a correction required  for accuracy in the atmospheric resonance region.    The Bessel function in Eq.~(\ref{ppluswasrr})  simplified by  using  its expanded representation given in Eq.~(\ref{Besssim}) below and  taking  $\Delta {\hat {\bar E}}[1] $ from Table~\ref{hatwAboveSolar}.  We find,
\beq
\label{ppluswpp1b}
P^{+ab}(\Delta_L,\hat A) &=&   \frac{ 4\Delta_L^2}{ {\hat {\bar D} } }  ( ( {\hat {\bar w}}_{0}^{ab} [1]  + D {\hat {\bar w}}_{0}^{e \mu} [1] )  j_0^2 ( \Delta_L  {\hat C}_\alpha )    +2 r[\ell]   {\hat {\bar w}}_{0}^{ab} [1]    (j_0( 2 \Delta_L  {\hat C}_\alpha )-  j_0^2 ( \Delta_L  {\hat C}_\alpha ) ) ) ~,
\eeq
where the term $ r[\ell]  D {\hat {\bar w}}_{0}^{e \mu} [1] $ has been dropped because it is of order $\alpha^3$ and therefore quite small.  The
 short-hand notation $\tilde A \equiv 1 -  {\hat A } $ is used in Table~\ref{hatwAboveSolar} and below for the frequently occurring quantity $1 -  {\hat A }$.

\subsection{ Simplifying $j_0^2( \Delta_L \Delta {\hat {\bar E}}[\ell]  ) $ }

The Bessel functions,
\beq
j_0 ( \hat{ \Delta}[\ell])    &\equiv&  j_0( \Delta_L \Delta {\hat {\bar E}}[\ell]  )  
\eeq
may be simplified when $\Delta {\hat {\bar E}}[\ell] $ consists of two leading terms,
\beq
\Delta {\hat {\bar E}}[\ell] &=& \Delta {\hat {\bar E}}_0 [\ell] ( 1 +r[\ell] ) ~,
\eeq
 as it does above the solar resonance region, and, in addition, $ \Delta_L r[\ell]  \Delta {\hat {\bar E}}_0 [\ell] <<1 $.  In this case,  the trigonometric identity,
\beq
&& \sin \Delta_L (  \Delta {\hat {\bar E}}_0 [\ell]   + r[\ell]  \Delta {\hat {\bar E}}_0 [\ell] ) \nonumber \\
&=&  \sin \Delta_L   \Delta {\hat {\bar E}}_0 [\ell] ~   \cos \Delta_L r[\ell]  \Delta {\hat {\bar E}}_0 [\ell]  \nonumber \\
&+&  \cos \Delta_L  \Delta {\hat {\bar E}}_0[\ell]  ~ \sin \Delta_L r[\ell]  \Delta {\hat {\bar E}}_0 [\ell] ~,
\eeq
and the  approximations,
\beq
 \cos \Delta_L r[\ell]  \Delta {\hat {\bar E}}_0 [\ell]  &\approx & 1 \nonumber \\
 \sin \Delta_L r[\ell]  \Delta {\hat {\bar E}}_0 [\ell]  &\approx& \Delta_L r[\ell]  \Delta {\hat {\bar E}}_0 [\ell] ~,
 \eeq 
allow us to write,
\beq
\label{trigiden0}
j_0( \Delta_L \Delta {\hat {\bar E}}[\ell]  )  &=& \frac{1}{ 1 +r[\ell]  }   j_0( \Delta_L \Delta {\hat {\bar E}}_0[\ell]  ) \nonumber \\
&+&  \frac{r[\ell] }{ 1 +r[\ell]  }  \cos \Delta_L  \Delta {\hat {\bar E}}_0[\ell]  ~ .
\eeq
In Eq.~(\ref{trigiden0}), 
\beq
j_0( \Delta_L \Delta {\hat {\bar E}}_0[\ell] )  &\equiv&  \frac{ \sin \Delta_L   \Delta {\hat {\bar E}}_0 [\ell]  }{\Delta_L  \Delta {\hat {\bar E}}_0 [\ell]  } ~.
\eeq

Squaring Eq.~(\ref{trigiden0}) and retaining the first two leading terms, we find

 \beq
j_0^2( \Delta_L \Delta {\hat {\bar E}}[\ell]  )  &=&  \frac{1}{ (1 +r[\ell] )^2  } j_0^2 ( \Delta_L \Delta {\hat {\bar E}}_0[\ell]  )   + \frac{ 2 r[\ell]}{ (1 +r[\ell]  )^2 } j_0( \Delta_L \Delta {\hat {\bar E}}_0[\ell] ) \cos \Delta_L  \Delta {\hat {\bar E}}_0[\ell] ~.
\eeq
The Bessel function may be simplified by expanding it to first order in $r[1]$, 
 \beq
 \label{Besssim}
j_0^2( \Delta_L \Delta {\hat {\bar E}}[\ell]  )   &=&   j_0^2 ( \Delta_L \Delta {\hat {\bar E}}_0[\ell]  )  (1 - 2 r[\ell]  )   +2 r[\ell]  j_0( 2 \Delta_L \Delta {\hat {\bar E}}_0[\ell]  )  \nonumber \\
&=&   j_0^2 ( \Delta_L \Delta {\hat {\bar E}}_0[\ell] )    +2 r[\ell]  (j_0( 2 \Delta_L \Delta {\hat {\bar E}}_0[\ell] -  j_0^2 ( \Delta_L \Delta {\hat {\bar E}}_0[\ell] ) ) ~.
 \eeq

\end{widetext}
\begin{acknowledgements}

We are indebted to Ernest Henley who introduced us to the subject of neutrino oscillations and made important contributions to the present paper. LSK thanks LANL group P-25  for its support.

\end{acknowledgements}

\end{document}